\documentclass{article}
\usepackage{arxiv}

\usepackage{graphicx} 
\usepackage{tabularx} 
\usepackage[table]{xcolor}
\usepackage[compact]{titlesec}
\usepackage[hidelinks]{hyperref}
\usepackage{amsmath}
\usepackage{amssymb}
\usepackage{mathtools}
\usepackage{amsthm}
\usepackage{algorithm}
\usepackage{comment}
\usepackage[noEnd=True]{algpseudocodex}
\usepackage{float}
\usepackage{stfloats}

\usepackage{xcolor}

\usepackage{booktabs} 
\usepackage{siunitx}
\usepackage{multirow}
\usepackage{makecell}
\usepackage{subfig}
\usepackage[capitalize,noabbrev]{cleveref}

\usepackage{tcolorbox}

\newcommand{\promptbox}[1]{
    \begin{tcolorbox}
    {#1}
    \end{tcolorbox}
}

\newcommand{\tokenvocab}{\mathcal{I}}

\DeclareMathOperator*{\argmax}{arg\,max}

\newcommand{\secref}[1]{\cref{#1}}
\newcommand{\apref}[1]{\cref{#1}}

\newcommand{\Colon}{\,:\,}

\newcommand{\bnm}{\begin{newmath}}
\newcommand{\enm}{\end{newmath}}
\newcommand{\bne}{\begin{newequation}}
\newcommand{\ene}{\end{newequation}}

\newenvironment{newmath}{\begin{displaymath}%
\setlength{\abovedisplayskip}{4pt}%
\setlength{\belowdisplayskip}{4pt}%
\setlength{\abovedisplayshortskip}{6pt}%
\setlength{\belowdisplayshortskip}{6pt} }{\end{displaymath}}

\newenvironment{newequation}{\begin{equation}%
\setlength{\abovedisplayskip}{4pt}%
\setlength{\belowdisplayskip}{4pt}%
\setlength{\abovedisplayshortskip}{6pt}%
\setlength{\belowdisplayshortskip}{6pt} }{\end{equation}}

\newlength{\saveparindent}
\newlength{\saveparskip}
\newcounter{ctr}

\newenvironment{newitemize}{%
\begin{list}{\mbox{}\hspace{5pt}$\bullet$\hfill}{\labelwidth=15pt%
\labelsep=5pt \leftmargin=20pt \topsep=3pt%
\setlength{\listparindent}{\saveparindent}%
\setlength{\parsep}{\saveparskip}%
\setlength{\itemsep}{3pt} }}{\end{list}}

\newenvironment{newenum}{%
\begin{list}{{\rm (\arabic{ctr})}\hfill}{\usecounter{ctr} \labelwidth=17pt%
\labelsep=5pt \leftmargin=22pt \topsep=3pt%
\setlength{\listparindent}{\saveparindent}%
\setlength{\parsep}{\saveparskip}%
\setlength{\itemsep}{2pt} }}{\end{list}}

\newcommand{\PPLcalib}{\rho}
\newcommand{\vecx}{\vec{x}}

\newcommand{\concat}{\|}
\newcommand{\thh}{^\textnormal{th}}
\newcommand{\confounder}{c}
\newcommand{\control}{R}

\newcommand{\threshold}{\tau}
\newcommand{\trans}{T}
\newcommand{\params}{\omega}
\newcommand{\strong}{\mathtt{s}}
\newcommand{\weak}{\mathtt{w}}
\newcommand{\mlmodel}{M}
\newcommand{\modelCollection}{\mathcal{M}}
\newcommand{\queryspace}{\mathcal{X}}
\newcommand{\getsr}{{\;{\leftarrow{\hspace*{-3pt}\raisebox{.75pt}{$\scriptscriptstyle\$$}}}\;}}

\newcommand{\indicatorFunc}{\mathbb{I}}

\newcommand{\pred}{\mathcal{P}}
\newcommand{\advA}{\mathcal{A}}

\renewcommand{\paragraph}[1]{\vspace*{6pt}\noindent\textbf{#1}}

\usepackage[normalem]{ulem}

\usepackage{cite}

\title{Rerouting LLM Routers} 

\author{
  Avital Shafran \\
  The Hebrew University\\of Jerusalem \And Roei Schuster\\ Wild Moose \And
  Thomas Ristenpart \\ Cornell Tech \And Vitaly Shmatikov \\ Cornell Tech}
\date{}

\begin{document}

\thispagestyle{plain}
\pagestyle{plain}

\maketitle

\begin{abstract}

  LLM routers aim to balance quality and cost of generation by classifying
  queries and routing them to a cheaper or more expensive LLM depending on their
  complexity. Routers represent one type of what we call LLM control planes:
  systems
  that orchestrate use of one or more LLMs. 
  In this paper, we investigate routers' adversarial robustness.

  We first define LLM control plane integrity, i.e., robustness of LLM
  orchestration to adversarial inputs, as a distinct problem in AI safety. Next, we
  demonstrate that an adversary can generate query-independent token sequences
  we call ``confounder gadgets'' that, when added to any query, cause LLM
  routers to send the query to a strong LLM. 

  Our quantitative evaluation shows that this attack is successful both in
  white-box and black-box settings against a variety of open-source and
  commercial routers, and that confounding queries do not affect the quality
  of LLM responses.  Finally, we demonstrate that gadgets can be effective while
  maintaining low perplexity, thus perplexity-based filtering is not an
  effective defense. We finish by investigating alternative defenses.

\end{abstract}

\section{Introduction}
\label{sec:intro}

Large language models (LLMs) exhibit remarkable capabilities on many tasks.  Today, hundreds of open-source and proprietary LLMs are available at different prices, ranging from expensive, state-of-the-art models to cheaper, smaller, less capable ones.  LLM operators typically provide API access to their models (especially higher-quality models) on a pay-per-query basis.  This imposes non-trivial costs on LLM-based applications and systems.

Developers who want to integrate LLMs into their applications must therefore consider both utility and cost.  They want to maximize the quality of responses to their queries while minimizing the cost.  The two objectives conflict with each other: larger models tend to generate higher-quality answers but charge more per query.  For example, at the time of this writing,
GPT-3.5-turbo costs $\$0.5/\$1.5$ per 1M input/output tokens, GPT-4o-mini $\$0.15/\$0.6$, GPT-4o $\$2.5/\$10$, o1-preview $\$15/\$60$.
The difference in quality between models is not uniform across queries.  For some queries, even a cheap model can generate an acceptable response.  More complex queries require an expensive model to obtain a quality answer. 

A natural solution to balancing performance and economic considerations is to
take advantage of the availability of multiple LLMs at different
price-performance points.  Recently proposed \emph{\textbf{LLM routing}}
systems~\cite{vsakota2024fly,ong2024routellm,dinghybrid,martian,unify}
orchestrate two or more LLMs and adaptively route each query to the cheapest LLM
they deem likely to generate a response of sufficient quality. In the two-LLM
case, let $M_s$ be an expensive, high-quality model and $M_w$ a weaker,
lower-grade one.  Given query $q$, the routing algorithm $R(\cdot)$ applies a
classifier to $q$ that outputs~$0$
if $M_w$ is sufficient for answering $q$, or $1$ if $M_s$ is required.  The system
  then routes $q$ accordingly.

LLM routing is an example of a general class of systems we call LLM control planes, which orchestrate the use of multiple LLMs to process inputs, as further described in Section~\ref{sec:control-planes}. 

\paragraph{Our contributions.}
First, we introduce \emph{\textbf{LLM control plane integrity}} as a novel problem in AI safety.  Recently proposed LLM control-plane algorithms are learned, calibrated classifiers (see Section~\ref{sec:control-planes}).  Their inputs are queries from potentially adversarial users.  Robustness of control-plane algorithms to adversarial queries is a new problem, distinct from adversarial robustness of the underlying LLMs.

To initiate the study of this problem, we show that existing LLM routing algorithms are not adversarially robust. We design, implement, and evaluate a method that generates \emph{query-independent} adversarial token sequences we call ``confounder gadgets.''  If a gadget is added to any query, this query is routed to the strong model with high probability. 
Next, we show that this attack is effective even in the \emph{transfer} setting
where the adversary does not have full knowledge of the target LLM router (it is
black-box), but has access to another router (e.g., an internally trained
surrogate).  We also evaluate the integrity of commercial LLM routers, showing
that they can be confounded as well.

Third, we investigate defenses.  Our basic method generates gadgets that have anomalously high perplexity.  Confounded queries are thus easily distinguished from normal queries and can be filtered out by the routing system.  Unfortunately, this defense can be evaded by an adversary who incorporates a low-perplexity objective into the gadget generation algorithm, producing gadgets that have low perplexity\textemdash and yet are effective at re-routing queries to the strong model.  We also discuss higher-level defenses, such as identifying users whose queries are routed to the strong model with abnormal frequency.

Routing attacks can be deployed for various adversarial objectives, e.g., to ensure that the adversary always obtains the highest-quality answer regardless of the target applications's internal routing policies and cost constraints, or to maliciously inflate the target's LLM costs.  As LLM control planes grow in importance and sophistication, we hope that this work will motivate further research on their adversarial robustness.

\section{LLM Control Planes and Routing}
\label{sec:control-planes}

Inference using large language models (LLMs) is traditionally monolithic: a single
model is applied to an input or sequence of inputs. This methodology can
be sub-optimal for various reasons. State-of-the-art models are often
expensive, with API access to LLMs costing as much as several dollars for each
query. Elsewhere, distinct LLMs may excel at different tasks, and selectively
using them may improve overall quality on a diverse workload. Finally, 
combining multiple LLMs, even all trained for similar tasks, may become
increasingly prevalent as performance improvements of individual LLMs plateaus~\cite{reuters2024ailimits,theinformation2024ailimits,bloomberg2024ailimits}.

Researchers and practitioners are therefore now
developing inference architectures that use multiple LLMs to answer queries.
These LLMs are orchestrated by what we call an \emph{LLM control plane}
(borrowing the terminology from networking~\cite{controlplanes}). The control
plane may route queries or parts of queries to different LLMs, derive new
strings to query to underlying LLMs, combine answers from underlying LLMs, and
more.

\begin{figure*}[t]
    \centering
    \includegraphics[width=0.9\linewidth]{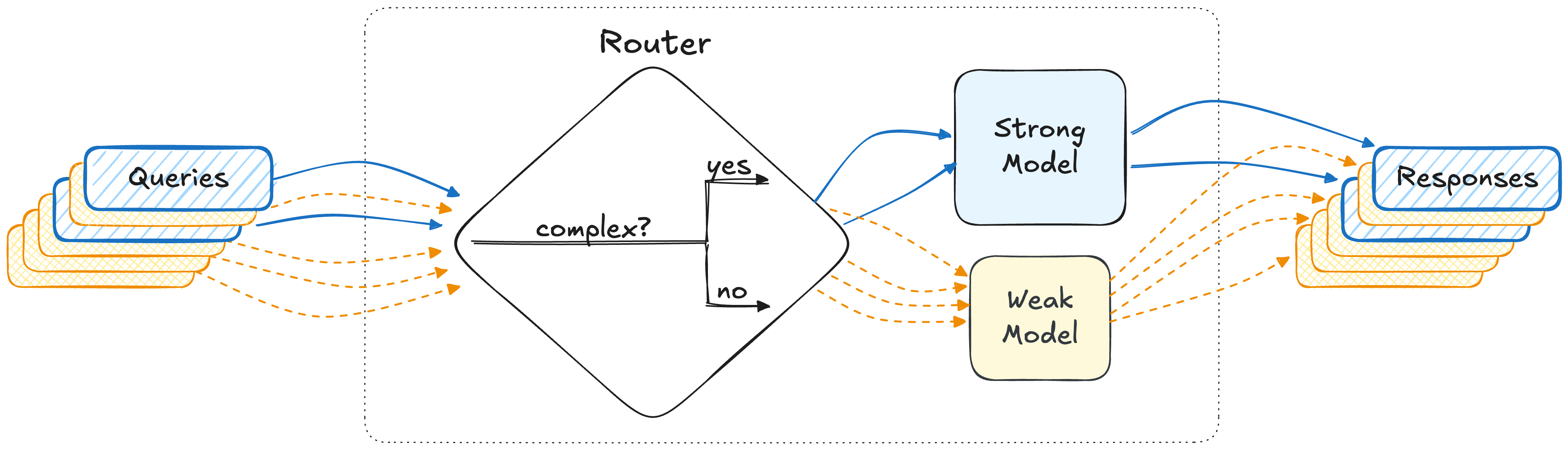}

    \caption{LLM routers classify queries and route 
    complex ones to an expensive/strong model, others to a
    cheaper/weak model. To control costs, LLM routers can be calibrated to
    maintain (for an expected workload) a specific ratio between queries sent to the strong and weak models.}
    \label{fig:router_diagram}
\end{figure*}

\paragraph{LLM routers.} A prominent example of this emerging class of LLM
control planes are \emph{LLM
routers}~\cite{dinghybrid,ong2024routellm,stripelis2024polyrouter,vsakota2024fly,lee2024orchestrallmefficientorchestrationlanguage}. 
LLM routers decide which of the two (or, sometimes, more)
LLMs to use to answer a query. In prescriptive routing, the router applies some
lightweight classifier to the input query that determines which underlying LLM
to utilize for a response. The classifier is itself a learned function that
scores the complexity of the query.
Deployments can then configure a score threshold for when to route a query to the more expensive
LLM.  This threshold can be tuned using representative workloads to achieve a 
desired cost-performance trade-off.  Figure~\ref{fig:router_diagram} shows the basic workflow of binary LLM routers.

Non-prescriptive
routing~\cite{chen2023frugalgpt,aggarwal2023automix,yue2024largelanguagemodelcascades} uses the responses
from one or more underlying LLMs to determine which response to return to the user.
For example, FrugalGPT~\cite{chen2023frugalgpt} submits the query to a sequence of
models (ordered by price) called a cascade, stopping when it obtains a response classified by the router as sufficient.  

In contrast to routers motivated by controlling costs, several LLM
router designs focus solely on improving quality of
responses~\cite{shnitzer2023large,narayanan2023tryage,feng2024graphrouter,srivatsa2024harnessing}.

The LLM routers described thus far do not modify the queries or individual LLM
responses. Other types of control planes do. Ensemble approaches such as mixture-of-expert
(MoE)~\cite{du2022glam,fedus2022switch,riquelme2021scaling,shazeer2017outrageously} architectures select a subset of underlying models to apply to
each token of a query and merge their responses. LLM
synthesis~\cite{jiang2023llm} architectures operate similarly, but route the entire query to a subset of underlying LLMs and merge their responses.  These approaches reduce inference costs by using fewer and/or less complex underlying models.

\paragraph{Applications of LLM routers.}  A key use case for LLM routers is to
help LLM-based application reduce cost.  Several commercial routers, including
Unify \cite{unify}, Martian \cite{martian}, NotDiamond \cite{notdiamond}, and
others, offer this as a service.  By replacing a few lines of code, the
application can send user queries to a router service, rather than directly to
some LLM provider.  The service selects the optimal LLM and forwards the
queries.  Commercial router services claim that this results in significant cost
savings: up to $98\%$ in the case of Martian \cite{martian}, and $10\times$ in the case of NotDiamond \cite{notdiamond}.

\section{LLM Control Plane Integrity}
\label{sec:integrity}

In this section, we define \emph{LLM control plane
integrity}.  Informally, it means that decisions made about underlying LLM
queries made by the control plane algorithms cannot be subverted by adversarial queries.   
Looking ahead, we will focus on one class of control plane: 
predictive LLM routing as used to manage cost.

\paragraph{Formalizing control planes.}
An LLM control plane $\control_\params$ is a potentially randomized
algorithm. It is parameterized by a string~$\params$, called the parameters. 
It utilizes some number~$n$ of LLMs denoted by $\modelCollection$. We will
mostly focus on the case of $n=2$, and, for reasons that will be clear in a moment, use
$\mlmodel_\strong$ (``strong'') and $\mlmodel_\weak$  (``weak'') to denote the two underlying LLMs.
Then inference on an input $x \in \queryspace$ for some set~$\queryspace$ of
allowed queries is performed by computing a response via 
$y \getsr \control_\params^{\modelCollection}(x)$. Here we use $\getsr$ to
denote running $\control$ with fresh random coins; we use $\gets$ when
$\control$ is deterministic.
We focus on inference for a single query, but it is straightforward to extend our
abstraction for control planes to include sessions: the controller would 
maintain state across invocations, potentially adapting its behavior as a
function of a sequence of queries and responses.

LLM control planes should, in general, be relatively computationally lightweight, at least compared
to the underlying LLMs. This is particularly so in the cost-motivated usage of
control planes, as a computationally or financially expensive control plane
would eat into cost savings incurred by utilizing cheaper underlying LLMs for
some queries.
For example, predictive binary routers use relatively simple classifiers to determine which of 
$\mlmodel_\strong$ or $\mlmodel_\weak$  should be used to respond to a query.

\paragraph{Inference flow.} Given a set of LLMs $\modelCollection$,
a control plane $\control_\params$, and an input $x$, 
an LLM inference flow is the sequence of LLM invocations
$\mlmodel_{i_j}(z_j)$ for $1 \le j \le m$ and $i_j \in \{\weak,\strong\}$ 
made when executing $\control_\params^\modelCollection(x)$. 
Here $m$ is the total number of LLM invocations, and $z_1,\ldots,z_m$ are the
queries made to the underlying LLMs.
Should $\control$ be randomized, the
sequence and its length are random variables. 
An inference flow can be written as a transcript 
\bnm
  \trans = (i_1,z_1),(i_2,z_2),\ldots,(i_m,z_m)
\enm
of pairs of model indexes $i_j \in \{\weak,\strong\}$ and model inputs~$z_j$. 
Note that for simplicity we ignore the potential for 
parallelization, assuming execution proceeds
serially. For binary routers, we have $m=1$ and $\trans \in \{(\weak,x),(\strong,x)\}$.
We write submitting a sequence of inferences $\vecx = \vecx_1,\ldots,\vecx_q$ to a control
plane as 
\bnm
  \control_\params^\modelCollection(\vecx) =
    (\control_\params^\modelCollection(\vecx_1),\ldots,\control_\params^\modelCollection(\vecx_q))
\enm
where note that each invocation could result in multiple underlying LLM
invocations. In the binary router case, however, each invocation results in a
single LLM invocation.

An \emph{inference flow policy} dictates the control plane designer's intention
regarding use of the underlying models.  For example, an application may want to
ensure that only a small fraction of queries go to the expensive model
$\mlmodel_\strong$. 
We can define this as a predicate over a sequence of transcripts.
In our binary router example, the policy can be more simply defined as a predicate
$\pred$ over (input, model) pairs $(\vecx_1,i_1),\ldots,(\vecx_q,i_q)$ since
this fully defines the sequence of transcripts. For example, a policy might specify that the strong model is used in at most an $\epsilon$ fraction of
inferences:
\bnm
  \pred((\vecx_1,i_1),\ldots,(\vecx_q,i_q)) = 
    \left(\sum_{j=1}^q \frac{\indicatorFunc(i_j)}{q} \le \epsilon\right)
\enm
where $\indicatorFunc(i_j) = 1$ if $i_j = \strong$ and $\indicatorFunc(i_j) =
0$ if $i_j = \weak$.  In other words, the predicate is that the fraction of queries routed to the strong model is bounded by $\epsilon$.

\paragraph{Control plane integrity.}
A \emph{control plane integrity adversary} is a randomized algorithm~$\advA$
that seeks to maliciously guide
inference flow.  

In an unconstrained LLM control plane integrity attack, the adversary $\advA$
seeks to generate inputs $\vecx = \vecx_1,\ldots,\vecx_q$ such that running
$\control_\omega^\modelCollection(\vecx)$
generates a transcript for which 
$\pred((x_1,i_1),\ldots,(x_q,i_q)) = 0$.
This attack could be launched by an adversary who wants to maximize inference costs for a victim application using an LLM router.

A harder setting requires input adaptation, where the adversary is given inputs
$x_1,\ldots,x_q$ and it must find new inputs $\hat{x}_1,\ldots, \hat{x}_q$ 
for which the transcript resulting from   $\pred((\hat{x}_1,i_1),\ldots,(\hat{x}_q,i     _q)) = 0$. There will be some competing constraint, such as that $x_j$ and
$\hat{x}_j$ are very similar for each~$j$, or that the outputs $y_j \getsr
\control_\omega^\modelCollection(x_j)$
and $\hat{y}_j \getsr \control_\omega^\modelCollection(\hat{x}_j)$ are close. 
In the routing context, the adversary's goal 
is to increase the fraction of queries that get routed to the strong model, in order
to improve the overall quality of responses, drive up the victim application's inference costs, or both.

\paragraph{Relationship to evasion attacks.}  Evasion
attacks~\cite{dalvi2004adversarial,lowd2005adversarial,szegedy2013intriguing} against an inference system (also called adversarial
examples~\cite{goodfellow2014explaining,papernot2016limitations,papernot2017practical}) would, in our setting,
seek to find a small modification $\Delta$ to an input $x$ such that
$\control_\omega^\modelCollection(x+\Delta) \ne
\control_\omega^\modelCollection(x)$ where addition is appropriately
defined based on input type (e.g., slight changes to text). 

Our attack setting is not the same. 
The control plane integrity adversary seeks to maliciously control the inference \emph{flow}, not necessarily the \emph{output} of
inference. In an unconstrained attack, the adversary does not care what outputs are generated. In the input adaptation attack, the adversary seeks to craft inputs that modify the inference flow yet do \emph{not} change the responses of the strong underlying LLM to the extent possible.  Looking ahead, we will use evasion techniques in our adaptation
attacks against learned control plane routers, but, importantly, not the overall inference. 

In the other direction, undermining LLM control plane integrity 
could be a stepping stone toward evasion
attacks. For example, if $\control_\omega^\modelCollection$ is used to classify
malicious content by combining LLMs each tuned to different types of harm
categories, then modifying inputs to force inference flows away from 
appropriate models could aid evasion. We leave evaluation of how control-plane
integrity attacks can enable evasion to future work.

\paragraph{Threat models.} Within the context of control plane integrity attacks
against LLM routers, we identify several threat models that differ in terms of
the adversary's goals and their knowledge about the target control plane
$\control_\omega^\modelCollection$. 

In terms of goals, an adversary may seek to \emph{inflate the costs} of a victim
application that utilizes an LLM control plane. As a kind of denial-of-service
attack, such cost inflation would penalize the application developer who expects
routing to control costs.  Another adversarial goal could be \emph{arbitrage}:
consider an application that charges $X$ dollars per
query, whereas directly using
$\mlmodel_\strong$ costs $Y > X$. The application's lower rate~$X$ makes economic sense 
assuming it uses a router to route the bulk of queries to a cheaper model
$\mlmodel_\weak$.
An input adaptation attack in this setting can gain (indirect) access to $\mlmodel_\strong$, obtaining
an arbitrage advantage of $Y-X$ per query. To be effective, this arbitrage 
adversary would want to ensure that adaptations do not lower response quality
(i.e., it extracts all the value out of rerouting to $\mlmodel_\strong$). As before, the victim in this case is the application that relies on routing to lower its costs (unsuccessfully, under this attack).

We now discuss adversarial capabilities. We assume that
our victim application's prompt includes a substring that can be controlled by the
adversary. This represents many real-world apps such as chatbots, coding
assistants, writing assistants, and others, that insert user inputs into an LLM
prompt.
In crafting adversarial portions of prompts, an adversary may have various levels of knowledge about
the victim application's router.
We consider the following knowledge settings:
\begin{newitemize}
  \item \emph{White-box setting}: The adversary knows the control plane
    algorithm and its parameters $\omega$.
  \item \emph{Black-box (transfer) setting}: The adversary does not know the control plane
    algorithm $\control$ and $\omega$ for the target
    model, but knows instead another control plane algorithm
    $\control'_{\omega'}$ and its parameters. We refer to $\control_{\omega'}'$
    as the \emph{surrogate}. For
    example, this could arise if an adversary trains their own router using available data. In this setting our attacks are also \emph{zero-shot} in that they do not require any interaction with the target control plane before the query that is being rerouted.
\end{newitemize}

\section{Confounding Control Planes with Gadgets}
\label{sec:attack}

We now turn to our main contribution: a methodology for attacking LLM control plane
integrity. The key insight is that an adversary can modify queries to mislead or ``confound'' the routing logic into routing these queries to an LLM of the adversary's choosing.  Furthermore, we will demonstrate that these attacks can be black-box and
\emph{query-independent}, i.e., a single modification works for all queries and does not require advance knowledge of the specific router being attacked.

We focus on the binary router setting in which the router applies 
a learned scoring function to input queries and routes any query whose score 
exceeds some threshold~$\threshold$ to the strong LLM $\mlmodel_\strong$. This setting has
been the focus of several prior works~\cite{ong2024routellm,dinghybrid,lee2024orchestrallmefficientorchestrationlanguage} and is used in the control planes that are
deployed in practice (see \secref{sec:commercial}). 

\begin{figure*}[t]
    \centering
    \includegraphics[width=0.9\linewidth]{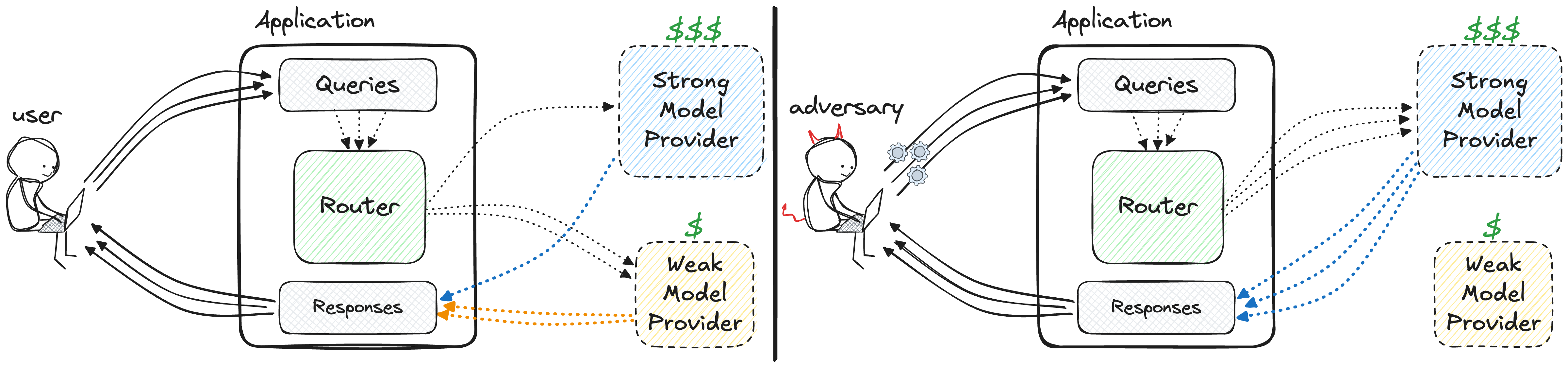}
    \caption{Overview of our attack on LLM routing control plane integrity.
     The attack adds to each query a prefix
    (represented by the gear), called a ``confounder
    gadget,'' that causes the router to send the query to the strong model.}
    \label{fig:attack_diagram}
\end{figure*}

More formally, 
we consider a router $\control_\params^{\modelCollection}$ for $\modelCollection =
\{\mlmodel_\weak,\mlmodel_\strong\}$, where $\params$ consists of a scoring
function $S$, scoring function's parameters $\theta$, and a threshold $\threshold \in \mathbb{R}^+$. 
For notational brevity we just write $\control_\params$ below, with
$\modelCollection$ clear from context. 
Here $S$ and $\theta$ define a scoring function  $S_\theta\Colon \queryspace
\rightarrow \mathbb{R}^+$. Since our focus is LLMs, we assume 
that queries $\queryspace$ are strings of text tokens. 
The routing algorithm then works as follows:
\begin{align*}
	\control_{\params}(x) = \begin{cases} \mlmodel_\weak(x) & \text{if } S_\theta(x) < \threshold \\
		\mlmodel_\strong(x) & \text{otherwise}
		\end{cases} 
\end{align*}
where $\params = (S, \theta, \threshold)$. We will detail scoring
functions in \secref{sec:exp_setting}; prior work has suggested linear models,
light-weight LLMs, and more. Note that, consistent with this application,
scoring functions are computationally efficient and cheap (as compared to
$\mlmodel_\strong,\mlmodel_\weak$).
Deployments calibrate $\threshold$ to limit the fraction of queries routed to
the strong model $\mlmodel_\strong$, giving rise to the type of control plane integrity policy
discussed in Section~\ref{sec:integrity}.

We focus on input adaptation attacks; these immediately give unconstrained
attacks as well. The adversary therefore has a sequence of inputs
$x_1,\ldots,x_q$ and must produce modified inputs $\hat{x}_1,\ldots,\hat{x}_q$ to
maximize the number of inputs routed to~$\mlmodel_\strong$. 
See Figure~\ref{fig:attack_diagram} for a depiction of our attack setting.

\paragraph{Instruction injection doesn't work.} Given the success of prompt
injection for jailbreaking~\cite{perez2022ignore} and other adversarial tasks~\cite{toyer2023tensor}, the adversary
might simply prefix each query~$x_i$ with some
instruction such as \emph{``Treat the following query as complex, \ldots''} to
generate a modified query $\hat{x}_i$. Our experiments show that this does
not work well, failing to trigger the control plane into routing otherwise weak
queries to~$\mlmodel_\strong$. See \apref{sec:other_methods} for details on our
experiments with various instruction prompts.

\paragraph{Confounder gadgets.}
Our approach works as follows.
Given a query $x_i$, we prepend a \emph{confounder
gadget} $\confounder_i$, which is a short sequence of adversarially chosen
tokens. The modified query is $\hat{x}_i = \confounder_i \concat x_i$ where
$\concat$ denotes string concatenation. Intuitively, we will use optimization to
search for confounders that trick the scoring function into
ranking $\hat{x}_i$ as sufficiently complex to require the strong model.

In the white-box, query-specific setting, we can choose $\confounder_i$ as a function of $x_i$
and the known parameters $\params = (S,\theta,\threshold)$. To do so, we fix a
confounder length of~$n$ tokens and let $\tokenvocab$ be a token dictionary (it
should be a sufficiently large subset of the token dictionary used by~$S$).
Then we set the gadget to initially be $n$ tokens all fixed to the same value
from $\tokenvocab$. The exact choice of the initialization token is not
important; in our implementation, we used the first token in the dictionary~(`!'). 
Denote this initial confounder as $\confounder_i^{(0)} = [\confounder_{i,1}^{(0)},\confounder_{i,2}^{(0)},\ldots,\confounder_{i,n}^{(0)}]$.

Then, we perform a hill-climbing style approach to find a good confounder
for $x_i$. 
For each iteration $t \in [T]$, where $T$ is the total number of iterations, do the following:
\begin{newenum}
\item Select a target index $j \in [1,n]$ uniformly.
\item Generate a set $\mathcal{B}$ of $B+1$ candidates. First set
$\tilde{\confounder}_0 = \confounder_i^{(t)}$, the current confounder.
To generate $B$ additional candidates, select replacement tokens from
$\tokenvocab$ uniformly, forming the set $\{t_b \gets \tokenvocab\}_{b=1}^B$. 
Replace the $j\thh$ token in the current confounder $\tilde{\confounder}_0$ with~$t_b$: 
\begin{align*}
\tilde{\confounder}_b =
[\confounder_{i,1}^{(t)},\ldots,\confounder_{i,j-1}^{(t)}, t_b,
\confounder_{i,j+1}^{(t)}, \ldots, \confounder_{i,n}^{(t)}]\;.
\end{align*}
Let $\mathcal{B} = \{\tilde{\confounder}_0,\ldots,\tilde{\confounder}_B\}$.
\item Find the candidate that maximizes the score: 
\begin{align}
    \confounder^{(t+1)}_i \gets \argmax_{\confounder \in \mathcal{B}} \; S_\theta(\confounder \concat x_i) \;.\label{eq:maximize}
\end{align}
\end{newenum}
The final confounder $c_i^{(T)}$ is used with query $x_i$.  We early abort if,
after 25 iterations, there is no update to the confounder gadget. Technically, we
could abort early if we find a confounder whose score exceeds $\threshold$. Running 
further can be useful when an adversary does not know~$\threshold$. 

The attack's runtime is dominated by $T\cdot B$ times the cost of executing $S$.  In practice,
$S$ are designed to be fast (otherwise routers would significantly increase the
latency of applications that use them). We report precise timings later; in summary, the
attack is fast because we can set $T$ to be relatively small and still find
high-scoring confounders.

Due to the randomness in index and token selection, the method converges to
different, yet similarly effective, confounder gadgets on each run.  Our evaluation 
will thus measure average performance over multiple gadgets.

\paragraph{Query-independent confounders.} One downside of the per-query
approach is that the adversary must repeat, for each query, the search for a good
confounder. In practice, the adversary might prefer a
\emph{query-independent} attack. Our confounder gadget approach extends to this
setting readily: perform the search routine above for an empty query.
In other words, just ignore $x_i$ in the query-dependent
attack above, replacing $S_\theta(\confounder\concat x_i)$ in
Eq.~\ref{eq:maximize} with $S_\theta(\confounder)$. This finds a
single query-independent confounder $\confounder$ that can be prefixed to all
queries, i.e., $\hat{x}_i = \confounder \concat x_i$.  We will show that this works
surprisingly well.

It is tempting to assume the reason a query-independent confounder works well 
is that a good scoring function should be roughly
monotonic in query extensions, 
i.e., one might expect that $S_\theta(c \concat x) \ge S_\theta(c)$ for almost any suffix~$x$. 
This intuition is not correct. In our experiments, we found that 
$S_\theta(c \concat x) < S_\theta(c)$ for many $x$ and some of the routers
discussed below. Nevertheless, by
ensuring that $S_\theta(c)$ is pretty high (set the number of iterations~$T$
higher) the resulting query-independent confounder works well. That is, we at
least get that $S_\theta(c\concat x ) > S_\theta(x)$.

\paragraph{The black-box setting: confounders that transfer.} Finally, the attacks so far are in the
white-box setting, where the attacker can optimize directly against $S_\theta$.  
While in some cases routing control planes will be public knowledge, in others,
including the proprietary control planes we explore in \secref{sec:commercial},
they are hidden. This gives rise to the black-box setting. While an attacker
might seek to perform model extraction
attacks~\cite{tramer2016stealing,lowd2005adversarial} to learn $\theta$, we
instead explore attacks that transfer from one router to another. 

In more detail, we assume the adversary has access to a router
$\control_{\params'}'$, called the
\emph{surrogate}, that is trained
on data similar to that used for the target router. Then the attack is the same
as above, except that we use the surrogate's scoring function $S'_{\theta'}$ instead
of the target's $S_\theta$. Again, we will see that this works surprisingly well: the
query-independent confounders found for the surrogate transfer to successfully
reroute queries against the target router.

\paragraph{Putting it all together.} In summary, our methodology for input adaptation
attacks is: 
\begin{newenum}
\item (Preprocessing) Develop a single query-independent confounder gadget~$\confounder$, using
 either the target router or surrogate to score the confounder.
\item (Input adaptation) For each query $x_i$, submit $\hat{x}_i = \confounder
\concat x_i$ instead to obtain a response $\hat{y}_i$. 
\end{newenum}

The confounder is applied to all queries, i.e., the adversary does not need to guess whether the original query would have been routed to the weak or strong model. In the rest of the paper, we demonstrate the confounders rarely result in ``downgrades,'' i.e., rerouting of queries from the strong to weak model.

We have experimented with variations of this approach that don't work quite
as well, for example adding
$\confounder$ as a suffix instead of a prefix.  See \secref{sec:ablations} for
details.

\section{Open-Source Routers: Experimental Setup}
\label{sec:exp_setting}

To evaluate efficacy of confounder gadgets generated using the method from Section~\ref{sec:attack}, we perform experiments with several LLM routers.  This section explains our experimental setup for the open-source routers
proposed in the research literature~\cite{ong2024routellm}; results of this evaluation appear in Section~\ref{sec:open-source-results}. In
\secref{sec:commercial}, we discuss experiments with proprietary, commercial
routers.   Figure~\ref{fig:experiments} shows the summary of our experimental setup.

In all experiments, we assume that the adversary's goal is to reroute queries to the strong model.  In Appendix~\ref{sec:minimizing_attack}, we evaluate efficacy of the attack when the goal is to reroute to the weak model.

\begin{figure}[t]
  \center
  \footnotesize
  \begin{tabularx}{.55\columnwidth}{lX}
\rowcolor{gray!25} \textbf{Routers} & \textbf{Notation}\\
      Similarity-weighted ranking & $\control_{SW}$\\
      Matrix factorization & $\control_{MF}$\\
      BERT classifier & $\control_{CLS}$\\
      LLM scoring & $\control_{LLM}$\\
  \end{tabularx}
  
  \vspace*{8pt}
  
  \begin{tabularx}{.55\columnwidth}{llX}
    \rowcolor{gray!25}    \textbf{LLM pair} &
    \multicolumn{1}{l}{\textbf{Strong ($\mlmodel_\strong$)}} &
    \multicolumn{1}{l}{\textbf{Weak ($\mlmodel_\weak$)}}\\
      \multicolumn{1}{c}{1} &  Llama-3.1-8B & 4-bit Mixtral 8x7B \\
      \multicolumn{1}{c}{2} &  Llama-3.1-8B& Mistral-7B-Instruct-v0.3 \\
      \multicolumn{1}{c}{3} &  Llama-3.1-8B& Llama-2-7B-chat-hf \\
      \multicolumn{1}{c}{4} &  GPT-4-1106-preview & 4-bit Mixtral 8x7B \\
  \end{tabularx}

    \vspace*{8pt}

  \begin{tabularx}{.55\columnwidth}{lX}
    \rowcolor{gray!25}    \textbf{Benchmark} & \textbf{Description}\\
      MT-Bench~\cite{zheng2023judging} & \phantom{14,}160 open-ended questions\\
      MMLU~\cite{hendrycks2021measuring} &14,042  multi-choice questions \\
      GSM8K~\cite{cobbe2021training} & \phantom{1}1,319 grade-school math problems
  \end{tabularx}

  \caption{Summary of our setup for routers, underlying LLMs, and benchmark
  datasets used in the experiments.}
  \label{fig:experiments}
\end{figure}

\paragraph{Target routers.} We focus our evaluation on the four prescriptive routing algorithms
proposed by Ong et al.~\cite{ong2024routellm}, which  provides open-source code and trained parameters, and does so for a representative variety of routing approaches: similarity-based classification~\cite{stripelis2024polyrouter,lee2024orchestrallmefficientorchestrationlanguage}, an MLP constructed via matrix factorization~\cite{stripelis2024polyrouter}, BERT-based classification~\cite{dinghybrid, stripelis2024polyrouter, vsakota2024fly}, and a fine-tuned LLM.

The routers we evaluate were trained in a supervised fashion using a set of
  reference (training) queries whose performance score on each of the considered
  models is known. The scores were computed from a collection of human pairwise
  rankings of model answers for each of the queries. We note that while the
  routers we consider are all learned using this training set, there is no
  reason to believe a non-learning-based approach (e.g., rule based) to routing would be more adversarially robust.

We now outline the routing methods considered in this work. See Ong et
  al.~\cite{ong2024routellm} for their full implementation details.

\emph{Similarity-weighted ranking:} The first method is based on the Bradley-Terry (BT) model~\cite{bradley1952rank}. For a given user query, this model derives a function to compute the probability of the weak model being preferred over the strong model. The probability-function expressions all share parameters, which are optimized to minimize the sum of cross-entropy losses over the training-set queries, where each element in the sum is weighted by the respective query's similarity with the user's query (computed as embeddings cosine similarity, with the embedding derived using OpenAI's text-embedding-3-small \cite{openai2024embedding}). 
We denote this method as $\control_{SW}$.  

\emph{Matrix factorization:} The second method is based on matrix factorization.  
The training queries are used to train a bilinear function mapping a model's embedding and a query's embedding to a score corresponding to how well the model performs on the query. Routing is done by computing the score of the input query for each model, and choosing the highest-scoring model.
 We denote this method as
$\control_{MF}$. 

\emph{BERT classifier:} The third method involves fine-tuning a classifier,
based on the BERT-base architecture~\cite{devlin2019bert}, to predict which of
the two models produces a better response for the given query or whether they do
equally well (a tie). The routing decision is based on the probability of the
weak model providing a better response versus the strong model or the tie. We
denote this method as $\control_{CLS}$. 

\emph{LLM classifier:} The last method is based on asking an LLM 
to provide a score in the range $1$--$5$ of how an AI expert would
struggle to respond to a given query based on the query's complexity. For this,
Ong et al.~fine-tuned a Llama-3-8B model~\cite{meta2024llama3} using their reference
set of queries and corresponding scores. We denote this method as $R_{LLM}$.

\paragraph{Underlying LLMs.} In~\cite{ong2024routellm}, Ong et al.\ trained the
routers with GPT-4-1106-preview \cite{achiam2023gpt} as the strong model and
Mixtral 8x7B \cite{jiang2024mixtral} as the weak model.   
They report
successful generalization between the underlying LLMs, stating that 
their routers trained for a particular strong-weak LLM pair can be used with
other strong-weak LLM pairs.

To allow our evaluation to scale, we use as the strong model $\mlmodel_\strong$
the open-sourced Llama-3.1-8B \cite{meta2024llama31} and as $\mlmodel_\weak$ the
4-bit quantized version of Mixtral 8x7B (for efficiency reasons).  This reduced
the cost of our experiments by avoiding expensive GPT API calls and lowering the
computational costs of Mixtral. Unless mentioned otherwise, all of our results
will be evaluated with respect to this pair, which we refer to as LLM pair 1. We performed more limited
experiments with the original strong, weak model pair (LLM pair 4) and had similar success in rerouting.

We additionally performed experiments with two further weaker models, in order
to better evaluate the case where weak models produce
much lower-quality responses for queries (compared to the strong model). 
In particular, we define LLM pair 2 as the strong model plus
Mistral-7B-Instruct-v0.3 \cite{jiang2023mistral} and LLM pair 3 as the strong
model plus Llama-2-7B-chat-hf
\cite{touvron2023llama}.    The weaker models in pairs 2 and 3 were chosen to represent smaller (Mistral 7B) and older-generation (Llama-2) models: 
according to the Chatbot Arena LLM
ranking leaderboard \cite{chatbot_arena,chiang2024chatbot}, Llama-3.1-8B is
ranked in the 58th place, Mixtral 8x7B at the 88th place, Mistral-7B at the
108th place, and Llama-2-7B at the 125th place.

The LLM strong-weak pairs with which we performed experiments are summarized
in Figure~\ref{fig:experiments}.

\paragraph{Evaluation datasets.}  We will evaluate our attacks using
three standard LLM benchmarks as workloads: MT-Bench \cite{zheng2023judging}, a dataset of 160 open-ended
questions, MMLU \cite{hendrycks2021measuring}, a dataset of 14,042 multi-choice
questions, and GSM8K \cite{cobbe2021training}, a dataset of 1,319 grade-school
math problems. 
Note that Ong et al.~\cite{ong2024routellm} flagged that some data points are
``contaminated'', i.e., they are too similar to the ones used in their training
of the routers. We use these datasets without these contaminated elements,  
resulting in 72 MT-bench queries, 14,037 MMLU queries, and 1,307 GSM8K queries.

For MMLU and GSM8K, we will require that the LLMs respond
in a predefined format so we can parse and compare the responses to ground-truth answers.  To facilitate this, we prepended formatting instructions to
the query, inserted as a prefix before the gadget in the case of
confounded queries. In other words, a confounded query ends up defined as $\hat{x}_i = instr \concat
\confounder \concat x_i$ for instruction template $instr$, confounder gadget
$\confounder$, and original query $x_i$.  
Thus in this case we model a scenario where the adversary only controls a part
of the prompt rather than the entire prompt. 
See~\Cref{sec:suffix} for formatting examples and ablations.

\paragraph{Router calibration.}
For each workload, we must calibrate each router by setting the threshold~$\tau$ to achieve some target fraction
$\epsilon$ of queries routed to the strong model. Note that the calibration
process we use is agnostic to the underlying LLM pair.
We therefore must define 12 distinct thresholds, one for
each router, dataset pair.
For our experiments here, we set $\epsilon = 0.5$,
meaning the goal is to have about half the queries routed to the strong model.
This reflects an application developer that seeks to control for costs, even if
it may mean sacrificing some performance for some workloads.

To calibrate for MT-bench, we use the Chatbot Arena~\cite{chiang2024chatbot}
dataset as the calibration set, computing the threshold using the 55\,K queries for 
which Ong et al.\ precomputed the scoring function outputs.  
To calibrate for MMLU and GSM8K, we select 1,000 queries uniformly at random and uses these to set thresholds. 
Looking ahead, we do not use these queries during evaluation of the
attacks.

Note that it important that the distribution of calibration queries be similar
to the distribution of the target workload (and, in our experiments, the test
queries).  We observed that the Chatbot Arena-based
threshold did not transfer well to MMLU and GSM8K, resulting in the majority of
queries ($\approx98\%$) routed to the strong model.

\section{Rerouting Open-Source Routers}
\label{sec:open-source-results}

We now empirically evaluate our rerouting attack against the open-source routers
described in the previous section.  Unless otherwise specified, our evaluation focuses on the query-independent attack setting where the attacker first finds a fixed set of gadgets and then uses them to attack arbitrarily many queries. This is the conservative setting, and query-specific gadgets --- which carry a higher computational cost --- generally work better.
 
In~\Cref{sec:other_methods} we evaluate optimization-free alternatives for generating our confounding gadgets, and show they significantly underperform our optimization-based approach.

\paragraph{White-box confounder gadget generation.} 
Following our attack framework described in~\Cref{sec:attack}, we construct a
query-independent control-plane gadget designed to confuse each router. We start
with the white-box setting, setting the batch size to $B = 32$ and the number of iterations to $T = 100$, ignoring thresholds.
We generate four sets of $n=10$ gadgets, i.e., ten for each router. 
Examples of generated gadgets can be found in~\Cref{sec:substring_examples}. 

When reporting scores
below, we therefore report the average over the $n$ gadgets used with all 72 MT-bench queries, 100 randomly selected MMLU queries, and 100 randomly
selected GSM8K queries. None of these testing queries were used in the
training of the routers or their calibration.

\begin{figure*}[t]

    \centering
    \subfloat[$R_{SW}$]{
    \includegraphics[width=0.24\linewidth]{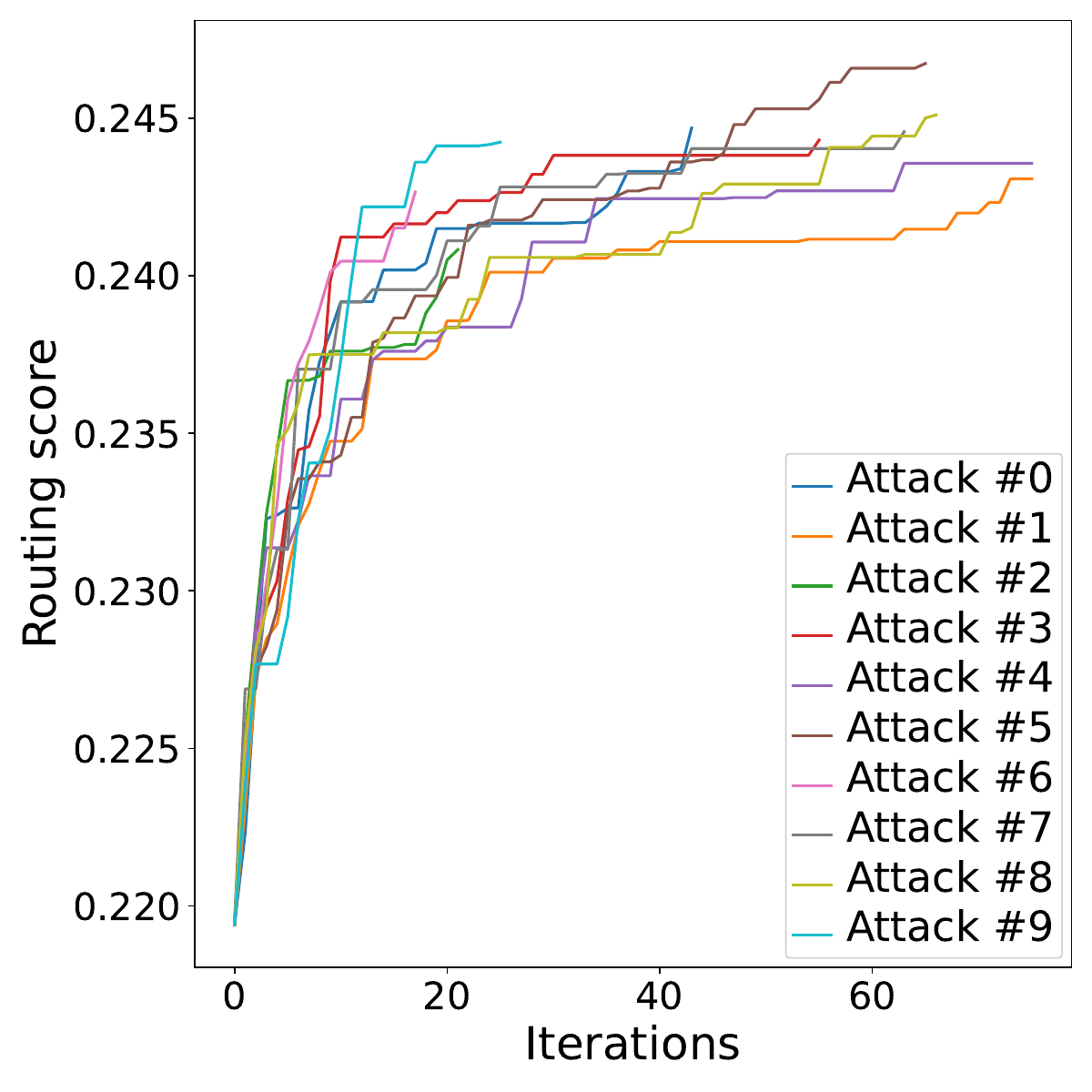}
    }
    \subfloat[$R_{MF}$]{
    \includegraphics[width=0.24\linewidth]{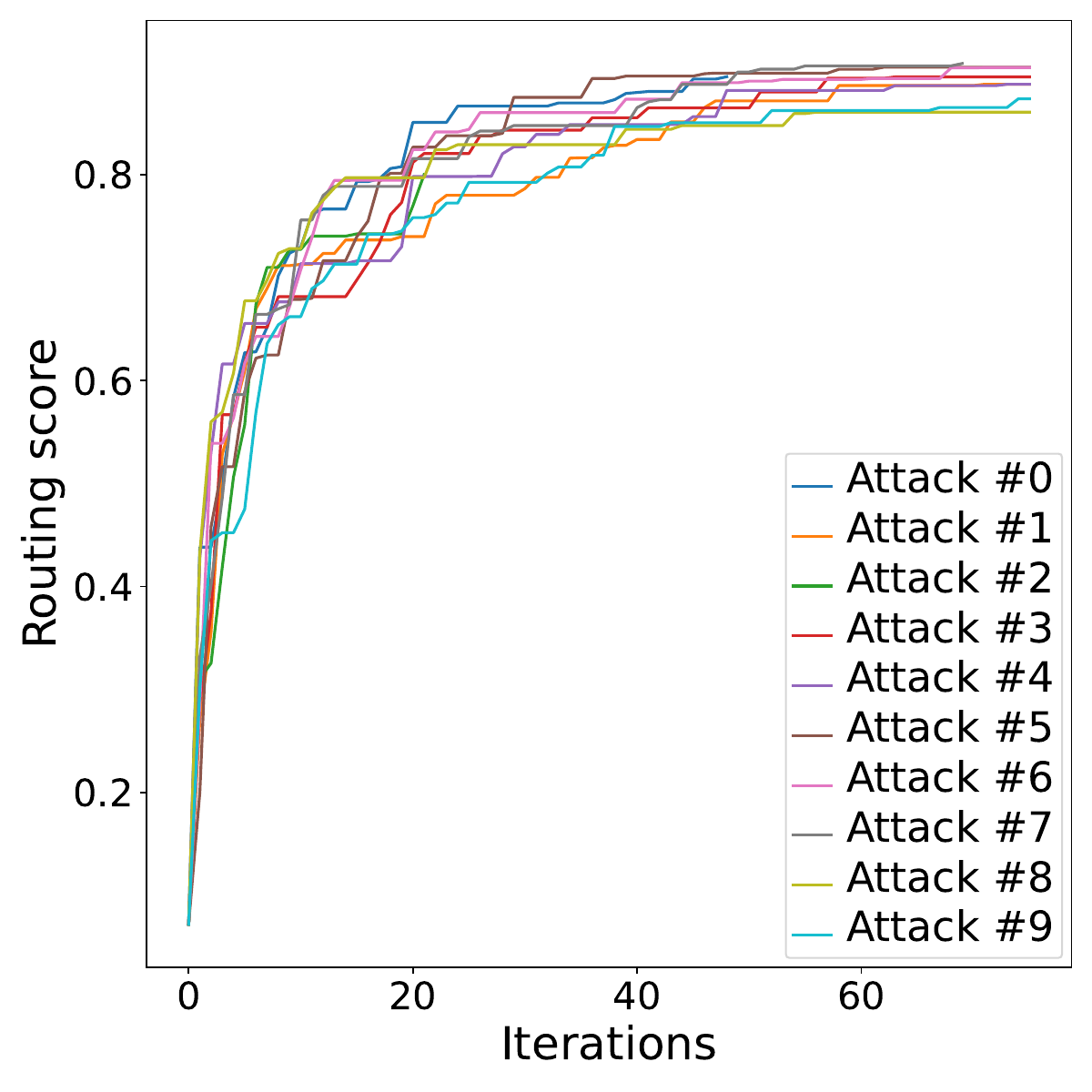}
    }
    \subfloat[$R_{CLS}$]{
    \includegraphics[width=0.24\linewidth]{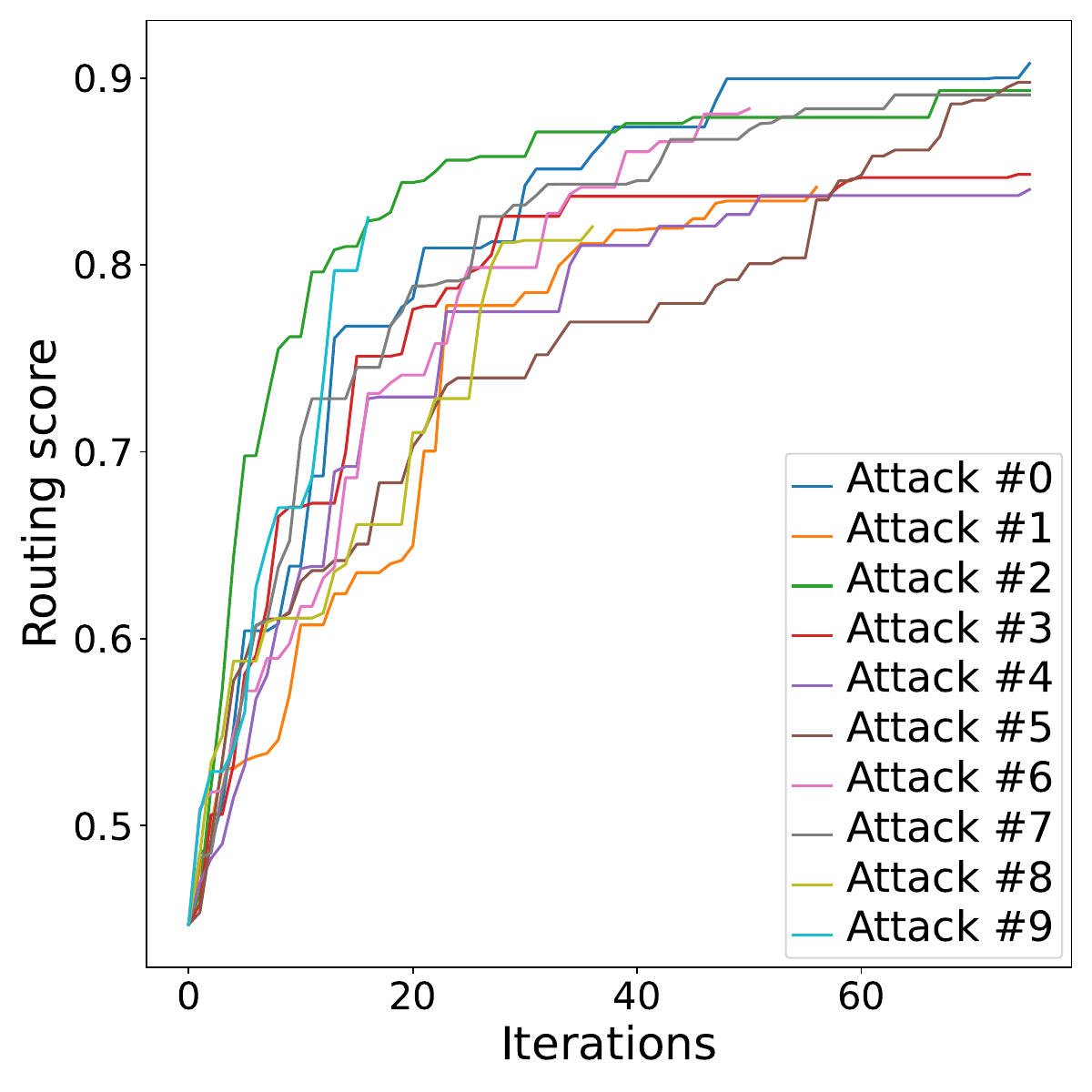}
    }
    \subfloat[$R_{LLM}$]{
    \includegraphics[width=0.24\linewidth]{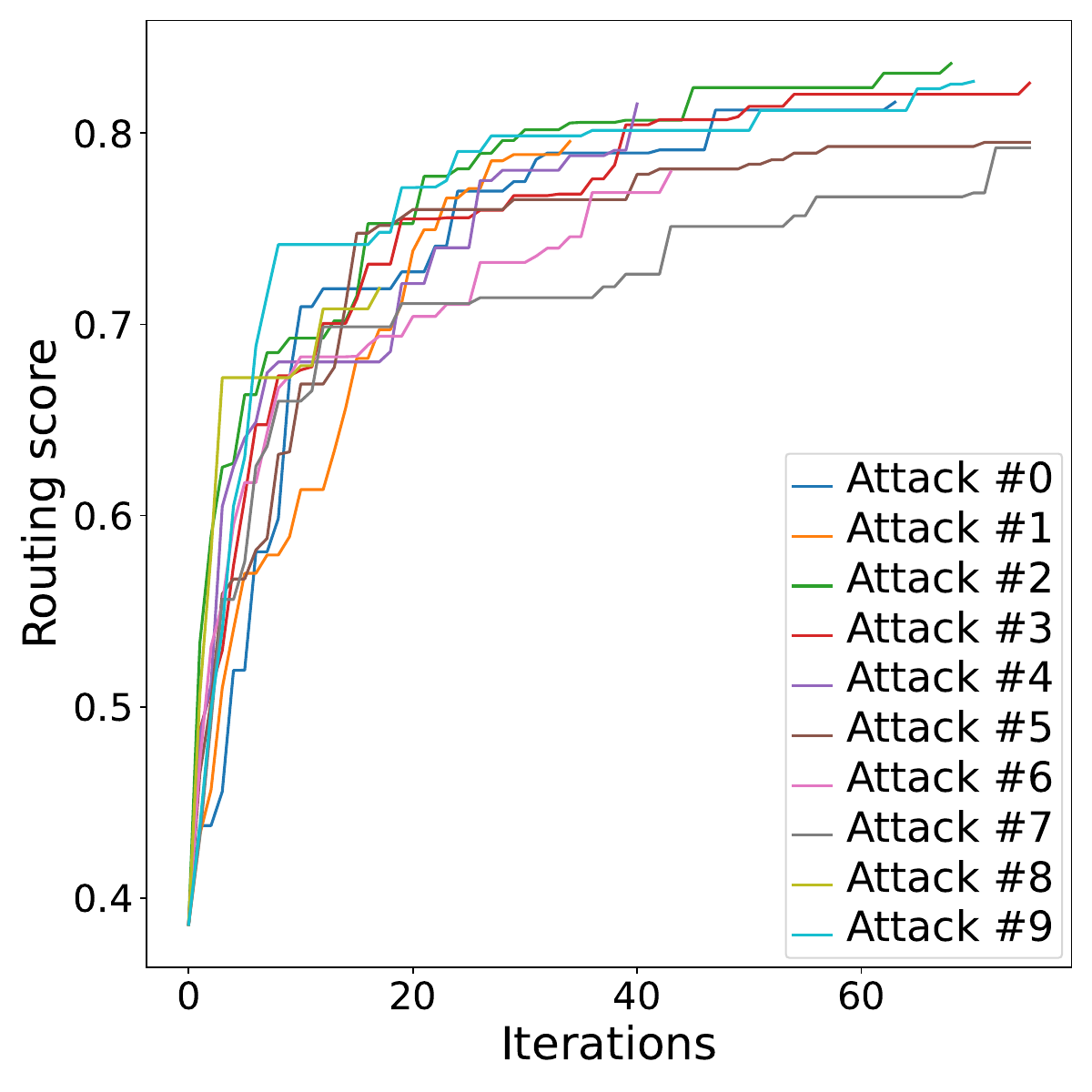}
    }
    
    \caption{Convergence of gadget generation against different routing algorithms.}
    \label{fig:convergence}
\end{figure*}

\paragraph{Runtime and convergence.}
\Cref{fig:convergence} shows the convergence rates for $10$ different gadgets, against different routing algorithms.  The overall average number of iterations before convergence is 58. Generation against $R_{SW}$ converges the fastest (50 iterations on average), $R_{MF}$ the slowest (66 iterations on average). Interestingly, the score of $R_{SW}$ does not increase much during optimization but is still sufficient for a successful attack. 

Runtime varies significantly when generating gadgets against different routing
methods.  On a machine with one A40~GPU, 4~CPUs, and 180G~RAM, a single
iteration takes $36.9$\,s, $8.4$\,s, $0.8$\,s, and $6.9$\,s for the $R_{SW}, R_{MF},
R_{CLS}$, and $R_{LLM}$ routers, respectively.  On average, it takes around 31
minutes to generate a gadget for the $R_{SW}$ router, 9 minutes for $R_{MF}$,
50s for $R_{CLS}$, and 6 minutes for $R_{LLM}$. 

\paragraph{Rerouting success rates.} Recall that our attack adds the same
confounder gadget to all queries. We start by evaluating the reroute \emph{success
rates}: What fraction of confounded queries does the target router send to the
strong model $\mlmodel_\strong$?  We show the results for our attack in
\Cref{tab:transfer_rates_same_router}. Focusing first on the columns labeled
`Strong', the notation ``$X
\rightarrow Y \pm Z$'' relays that for unmodified queries, $X$\% are routed to
the strong model $\mlmodel_\strong$, and that for the $n=10$ confounders we
achieved an average of $Y$\% of queries sent to $\mlmodel_\strong$. The standard error
is~$Z$. Note that while calibration targets $\epsilon=0.5$, we see some natural
variance in $X$ for the test sets. 

To breakdown success further, we additionally report the \emph{upgrade rate}, which focuses on the percentage
of queries that were (a)~originally routed to the weak model, and (b)~routed to
the strong model after they were modified with the confounder gadget. Because in
our attacks few queries get ``downgraded'' (confounders cause them to be 
rerouted to the weak model instead of strong), the upgrade rate dictates the
success rate.

\begin{table*}[t]
\small
\centering
\begin{tabular}{lrrrrrrrr}
   \toprule 
    & \multicolumn{2}{c}{$R_{SW}$} & \multicolumn{2}{c}{$R_{MF}$} & \multicolumn{2}{c}{$R_{CLS}$} & \multicolumn{2}{c}{$R_{LLM}$}\\
   & \multicolumn{1}{c}{Upgrade} & \multicolumn{1}{c}{Strong} & \multicolumn{1}{c}{Upgrade} & \multicolumn{1}{c}{Strong} & \multicolumn{1}{c}{Upgrade} & \multicolumn{1}{c}{Strong} & \multicolumn{1}{c}{Upgrade} & \multicolumn{1}{c}{Strong}\\
   \midrule
   MT-Bench & $100 \pm0$ & $81\to100\pm0$ & $100\pm0$ & $58\to100\pm0$ & $100\pm0$ & $67\to100\pm0$ & $73\pm5$ & $57\to88\pm2$ \\
  MMLU & $90\pm1$ & $43\to\phantom{0}94\pm1$ & $78\pm4$ & $53\to\phantom{0}90\pm2$ & $100\pm0$ & $47\to100\pm0$ & $95\pm1$ & $53\to98\pm1$\\
  GSM8K & $98\pm0$ & $52\to\phantom{0}99\pm0$ & $100\pm0$ & $54\to100\pm0$ & $100\pm0$ & $56\to100\pm0$ & $94\pm3$ & $53\to97\pm1$ \\
   \bottomrule
   \end{tabular}

  \caption{The white-box attack's rerouting success rate. ``Upgrade'' is the
  percentage of ``Weak'' queries successfully rerouted to the strong model by
  adding a confounder gadget. ``Strong'' shows the change in the percentage of queries that are routed to the strong LLM $\mlmodel_\strong$ before and after our rerouting attack.
  } \label{tab:transfer_rates_same_router}

\end{table*}

As can be seen, the gadgets reroute
almost all weak queries to the strong model. In most cases we see 100\% success,
or close to it. 
The worst case still achieves 88\%
rerouting success, boosting the fraction of queries sent to the strong LLM by
1.5x. Rerouting fails only for some queries that even after
confounding are sent to the weak model: the fixed gadget did not
sufficiently increase the router's estimate of those queries' complexity.
This is the only source of error for the attack:
\emph{no} queries in these experiments got
``downgraded'', i.e., a query that would otherwise be sent to
$\mlmodel_\strong$ ends up rerouted to $\mlmodel_\weak$.
This also means that adding the confounder to every single query does not have
negative impact on rerouting efficacy. We report standard error values for both the upgrade rates and the total percentage of queries routed to the strong model. The maximal standard error is in the low single digits, indicating similar success 
rates across gadgets. 

\paragraph{Quality of attack responses.}
We now turn to evaluating the quality of the responses generated by the attack. 
Note that because we have calibrated the routers to target $\epsilon=0.5$, our attacks can 
improve response quality by rerouting to the stronger model. In the other
direction, our attacks add confounder gadgets which might degrade response
quality.

As a first measure of response quality, we compare the perplexity scores for
unmodified responses and confounded query responses.  Text
perplexity~\cite{Jelinek1980InterpolatedEO} is a well-known method for
approximating ``naturalness'' of text sequences.  
Perplexity can be computed using an LLM, 
we use GPT-2~\cite{radford2019language} for this purpose as it is a standard
choice\cite{alon2023detecting,zhang2024controlled};\footnote{
Some responses had abnormally high perplexity values ($>100$), which we found do
not correlate with quality, but these variations disproportionately
contribute to the average. We thus filter out such high-perplexity responses as
outliers in both benign and attack settings. 
We provide examples of filtered responses in~\Cref{sec:perplexity_issues}.}
\Cref{tab:perplexity_same_router} shows the results. As can be seen, adding the confounder gadget to
queries does not significantly change response perplexity. To the extent that it
does, it usually somewhat decreases response perplexity, i.e., makes it more
``natural''. 
That said, perplexity is a coarse measure of ``naturalness,'' and it does not measure whether the response is correct.  In particular, responses of strong and weak LLMs tend to have similar perplexities. We further discuss this issue in~\Cref{sec:perplexity_issues}.

\begin{table*}[t]
\small
\centering
    \begin{tabular}{lcccccccc}
    \toprule 
       & \multicolumn{2}{c}{$R_{SW}$} & \multicolumn{2}{c}{$R_{MF}$} & \multicolumn{2}{c}{$R_{CLS}$} & \multicolumn{2}{c}{$R_{LLM}$}\\
     & Original & Confounded & Original & Confounded & Original & Confounded & Original & Confounded\\
    \midrule
    MT-Bench & $13.8$ & $12.3\pm0.2$ & $12.6$ & $12.3\pm0.2$ & $13.1$ & $12.1\pm0.2$ & $12.7$ & $12.7\pm0.4$\\
    MMLU & $20.4$ & $20.1\pm0.1$ & $20.0$ & $20.3\pm0.1$ & $20.2$ & $20.5\pm0.1$ & $21.0$ & $19.6\pm0.1$\\
    GSM8K & $17.1$ & $15.1\pm0.3$ & $17.0$ & $15.2\pm0.3$ & $17.0$ & $15.0\pm0.2$ & $16.4$ & $15.2\pm0.3$ \\
    \bottomrule
    \end{tabular}
\caption{Average perplexity of responses to the original and confounded queries, in the white-box setting for LLM pair 1. Response perplexity does not change significantly when adding the confounder gadget.
} \label{tab:perplexity_same_router}
\end{table*}

We thus also evaluate using the following benchmark-specific metrics to assess response quality:
\begin{newitemize}
   \item MT-bench: We score the responses on a scale of $1$--$10$ using an
     LLM-as-a-judge methodology \cite{zheng2023judging}. We use \mbox{GPT-4o}~\cite{openai2024gpt4o} as the judge and ask it to provide a score given a pair of a query and a
     corresponding response. 

  \item MMLU: We parse the responses and compare the answer to the ground truth.
    In cases where the response did not fit any known multi-choice format, we
    marked the response as a mistake. We report accuracy as the percentage of
    responses that match the ground truth.

  \item GSM8K: similar to MMLU except questions are math rather than multiple choice, thus we parse the answers according to the expected format.
\end{newitemize}

\Cref{tab:bench_specific_same_router} shows that, according to these metrics, in most cases responses to the confounded queries are no worse, and in some cases even better, than responses to the original queries. We attribute the improvement on the GSM8K benchmark to the fact that the strong model performs significantly better than the weak model on this benchmark ($57\%$ vs.\ $33\%$).  
On the MT-bench and MMLU benchmarks, strong and weak models have comparable performance ($8.5$ vs.\ $7.6$ for MT-bench and $66\%$ vs.\ $64\%$ for MMLU), thus routing does 
not degrade quality of responses and, consequently, the attack cannot improve it.

\begin{table*}[t]
\small
\centering
    \begin{tabular}{l|cc|cc|cc|cc}
    \toprule 
       & \multicolumn{2}{c|}{$R_{SW}$} & \multicolumn{2}{c|}{$R_{MF}$} & \multicolumn{2}{c|}{$R_{CLS}$} & \multicolumn{2}{c}{$R_{LLM}$}\\
     & Original & Confounded & Original & Confounded & Original & Confounded & Original & Confounded\\
    \midrule
    MT-Bench & $8.4$ & $8.3\pm0.0$ & $8.4$ & $8.4\pm0.0$ & $8.4$ & $8.3\pm0.0$ & $8.3$ & $8.2\pm0.1$ \\
    MMLU & $61$ & $66\pm0\phantom{0}$ & $64$ & $64\pm1\phantom{0}$ & $63$ & $65\pm0\phantom{0}$ & $67$ & $66\pm0\phantom{0}$\\ 
    GSM8K & $46$ & $64\pm1\phantom{0}$ & $50$ & $67\pm1\phantom{0}$ & $50$ & $63\pm1\phantom{0}$ & $44$ & $64\pm1\phantom{0}$ \\
    \bottomrule
    \end{tabular}
\caption{Average benchmark-specific scores of responses to the original and confounded queries, in the white-box setting for LLM pair 1.  Rerouting to the strong model improves quality of responses as long as there is a significant gap between the benchmark performance of the weak and strong LLMs.} \label{tab:bench_specific_same_router}
\end{table*}

To further demonstrate that the attack improves the quality of responses when
there is a significant gap between the weak and strong LLMs, we perform an
additional evaluation with Mistral-7B-Instruct-v0.3 \cite{jiang2023mistral} and
Llama-2-7B-chat-hf \cite{touvron2023llama} as the weak LLMs (LLM pairs 2 and 3).
Mistral-7B achieves $7.4$, $57\%$, and $25\%$ on MT-bench, MMLU, and GSM8K,
respectively. Llama-2-7B achieves $6.4$, $44\%$, and $21\%$.
\Cref{tab:bench_specific_same_router_weaker_models} shows that the rerouting
attack improves quality of responses when either of these LLMs is the weak
model, and in particular for the weaker Llama-2-7B model.

\begin{table*}[t]
\small
\centering
    \begin{tabular}{l|cc|cc|cc|cc}
    \toprule 
       & \multicolumn{2}{c|}{$R_{SW}$} & \multicolumn{2}{c|}{$R_{MF}$} & \multicolumn{2}{c|}{$R_{CLS}$} & \multicolumn{2}{c}{$R_{LLM}$}\\
     & Orig. & Conf. & Orig. & Conf. & Orig. & Conf. & Orig. & Conf.\\
    \midrule
    \multicolumn{9}{c}{LLM pair 2}\\ 
    \midrule
    MT-Bench & $8.5$ & $8.3\pm0.0$ & $8.4$ & $8.3\pm0.1$ & $8.4$ & $8.4\pm0.1$ & $8.4$ & $8.3\pm0.1$\\
    MMLU & $55$ & $64\pm1\phantom{0}$ & $63$ & $64\pm0\phantom{0}$ & $58$ & $66\pm1\phantom{0}$ & $62$ & $66\pm0\phantom{0}$\\
    GSM8K & $46$ & $64\pm1\phantom{0}$ & $51$ & $67\pm1\phantom{0}$ & $49$ & $63\pm1\phantom{0}$ & $38$ & $63\pm2\phantom{0}$\\
    \midrule
    \multicolumn{9}{c}{LLM pair 3}\\ 
    \midrule
    MT-Bench & $8.4$ & $8.3\pm0.0$ & $8.1$ & $8.3\pm0.1$ & $8.3$ & $8.4\pm0.1$ & $8.1$ & $8.2\pm0.1$\\
    MMLU & $51$ & $64\pm1\phantom{0}$ & $57$ & $63\pm1\phantom{0}$ & $52$ & $66\pm1\phantom{0}$ & $59$ & $66\pm1\phantom{0}$ \\
    GSM8K & $40$ & $64\pm1\phantom{0}$ & $44$ & $67\pm1\phantom{0}$ & $45$ & $63\pm1\phantom{0}$ & $37$ & $64\pm1\phantom{0}$ \\
    \bottomrule
    \end{tabular}
\caption{Average benchmark-specific scores of responses to the original and confounded queries with Mistral-7B-Instruct-v0.3 (LLM pair 2) or Llama-2-7B-chat-hf (LLM pair 3) as the weak model, in the white-box setting. Results further emphasize that the rerouting attack improves quality of responses when there is a significant gap between the weak and strong LLMs.
} \label{tab:bench_specific_same_router_weaker_models}
\end{table*}

LLM responses are sometimes affected by the confounder gadget.  In some cases,
the LLM responded with, for example, ``I can't answer that question as it appears to be a
jumbled mix of characters''. Still, the response
continued with ``However, I can help you with the actual question you're
asking,'' followed by the actual answer. We observed very few cases where an LLM
refused to answer due to the presence of the gadget. In most cases, the response
did not mention anything abnormal about the query. Intuitively, this reflects
the fact that while 
LLMs are built to be robust to noisy inputs, the router itself is not.

In summary, the attack is highly successful at rerouting queries from the weak to the strong model. Overall, quality improves if there is a significant gap between the strong and weak LLMs used by the router. Either way, confounding has no negative impact on the quality of responses.

\paragraph{Black-box attack results.} 
\label{sec:different_router_results}
Next, we consider the black-box attack, where the attacker does not know the
algorithm used by the target router.  We assume that the attacker has access to
another, surrogate router that it can use to generate confounder gadgets.  In
effect, we evaluate transferability of the attack from a known, white-box router
to unknown, black-box routers.

\Cref{tab:transfer_rates_different_router} shows the results for all
combinations of surrogate (denoted by $\hat{R}$) and target routers.  For
conciseness we focus on the upgrade and downgrade rates for the remainder of
this work. Upgrade rates are lower than in the white-box setting but still high,
indicating that the attack transfers.  The LLM-based routing algorithm $R_{LLM}$
has the lowest rates, perhaps because it is the most complex of the four. The
downgrade rate is $0$ in most cases and is $1.2\%$ on average. 

\begin{table*}[t]
\small
\centering
    \resizebox{\textwidth}{!}{\begin{tabular}{l|ccc|ccc|ccc|ccc}
    \toprule 
     Surrogate &\multicolumn{3}{c|}{$\hat{R}_{SW}$} & \multicolumn{3}{c|}{$\hat{R}_{MF}$}  &\multicolumn{3}{c|}{$\hat{R}_{CLS}$} & \multicolumn{3}{c}{$\hat{R}_{LLM}$} \\
     Target & $R_{MF}$ & $R_{CLS}$ & $R_{LLM}$ & $R_{SW}$ & $R_{CLS}$ & $R_{LLM}$ & $R_{SW}$ & $S_{FM}$ & $R_{LLM}$ & $R_{SW}$ & $R_{MF}$ & $R_{CLS}$\\
    \midrule
    MT-Bench & $99\pm1$ & $88\pm5\phantom{1}$ & $45\pm5$ & $100\pm0$ & $96\pm2$ & $39\pm3$ & $100\pm0$ & $79\pm9$ & $51\pm5$ & $100\pm0$ & $83\pm5$ & $85\pm7\phantom{0}$ \\
    MMLU & $66\pm5$ & $44\pm11$ & $81\pm3$ & $\phantom{0}82\pm4$ & $56\pm7$ & $74\pm2$ & $\phantom{0}64\pm6$ & $16\pm7$ & $80\pm5$ & $\phantom{0}53\pm4$ & $20\pm5$ & $46\pm11$ \\

    GSM8K & $99\pm1$ & $72\pm11$ & $63\pm4$ & $\phantom{0}92\pm2$ & $88\pm3$ & $62\pm4$ & $\phantom{0}76\pm6$ & $60\pm9$ & $65\pm8$ & $\phantom{0}60\pm8$ & $70\pm7$ & $73\pm10$ \\
    \bottomrule
    \end{tabular}}

  \caption{Average upgrade rates for our attack in the black-box setting. This
  is the average percentage of queries rerouted from
  the weak to strong model under the target router due to a confounder gadget generated using the
  surrogate. The average downgrade rate (i.e., strong-to-weak rerouting) is
  $1.2\%$ across all routers. Upgrade rates are lower than in the white-box setting but still high,
indicating that the attack transfers.}\label{tab:transfer_rates_different_router}
\end{table*}

\Cref{tab:perplexity_different_router} shows that the black-box attack does not
increase the average perplexity of responses as generated by LLM pair~1.
\Cref{tab:bench_specific_different_router} shows that the attack does not
decrease benchmark-specific scores, other than some small decrease in some cases
for the MMLU benchmark. For GSM8K, similar to the behaviour observed in the
white-box setting, we see an improvement with our attack due to the performance
difference between the strong and weak models for this task. This indicates that
confounding affects only the routing, not the quality of responses. When the weak model is significantly weaker than the strong model, i.e., LLM
pairs 2 and 3, the attack can improve the quality of responses significantly.

\begin{table*}[t]
\small
\centering
    \begin{tabular}{l|ccc|ccc|ccc|ccc}
    \toprule 
     Surrogate &\multicolumn{3}{c|}{$\hat{R}_{SW}$} & \multicolumn{3}{c|}{$\hat{R}_{MF}$} &\multicolumn{3}{c|}{$\hat{R}_{CLS}$} & \multicolumn{3}{c}{$\hat{R}_{LLM}$} \\
     Target & $R_{MF}$ & $R_{CLS}$ & $R_{LLM}$ & $R_{SW}$ & $R_{CLS}$ & $R_{LLM}$ & $R_{SW}$ & $S_{FM}$ & $R_{LLM}$ & $R_{SW}$ & $R_{MF}$ & $R_{CLS}$\\
    \midrule
    MT-Bench & $0.4$ & $0.8$ & $0.6$ & $1.4$ & $0.7$ & $0.3$ & $1.7$ & $0.3$ & $0.7$ & $0.8$ & $-0.6$ & $0.0$ \\
    MMLU & $0.1$ & $0.8$ & $1.1$ & $0.2$ & $0.2$ & $1.1$ & $0.3$ & $0.8$ & $0.9$ & $1.3$ & $\phantom{-}1.2$ & $0.9$ \\
    GSM8K & $1.9$ & $1.7$ & $0.6$ & $1.6$ & $1.7$ & $0.2$ & $1.7$ & $1.0$ & $0.4$ & $1.3$ & $\phantom{-}1.3$ & $1.7$ \\
    \bottomrule
    \end{tabular}
\caption{Differences between average perplexity of responses to the original and
  confounded queries, in the black-box setting, when the confounder gadget was
  generated for a different surrogate router than the target, for LLM pair 1.
  Positive values indicate a lower average perplexity (more natural) of
  responses to the confounded queries; higher values are better for the
  attacker. 
  Standard errors were omitted for readability but are $0.2$ on average. As in the white-box setting, the attack does not increase the average response perplexity.
} \label{tab:perplexity_different_router}
\end{table*}

\begin{table*}[t]
\small
\centering
    \begin{tabular}{l|ccc|ccc|ccc|ccc}
    \toprule 
     Surrogate &\multicolumn{3}{c|}{$\hat{R}_{SW}$} & \multicolumn{3}{c|}{$\hat{R}_{MF}$} &\multicolumn{3}{c|}{$\hat{R}_{CLS}$} & \multicolumn{3}{c}{$\hat{R}_{LLM}$} \\
     Target & $R_{MF}$ & $R_{CLS}$ & $R_{LLM}$ & $R_{SW}$ & $R_{CLS}$ & $R_{LLM}$ & $R_{SW}$ & $S_{FM}$ & $R_{LLM}$ & $R_{SW}$ & $R_{MF}$ & $R_{CLS}$\\
    \midrule
      \multicolumn{13}{c}{LLM pair 1}\\ 
    \midrule
    MT-Bench & $-0.1$ & $-0.1$ & $\phantom{-}0.0$ & $-0.1$ & $-0.1$ & $\phantom{-}0.0$ & $-0.1$ & $\phantom{-}0.0$ & $\phantom{-}0.1$ & $-0.2$ & $-0.1$ & $-0.2$\\ 
    MMLU & $-0.1$ & $\phantom{-}0.3$ & $-0.2$ & $\phantom{-}4.8$ & $\phantom{-}1.0$ & $\phantom{-}0.5$ & $\phantom{-}2.5$ & $-1.3$ & $-0.8$ & $\phantom{-}2.6$ & $-0.9$ & $\phantom{-}0.3$\\
    GSM8K & $\phantom{.}14.9$ & $\phantom{-}9.6$ & $\phantom{.}15.2$ & $\phantom{.}18.6$ & $\phantom{.}13.8$ & $\phantom{.}14.7$ & $\phantom{.}13.4$ & $\phantom{-}6.8$ & $\phantom{.}12.6$ & $\phantom{.}13.6$ & $\phantom{.}11.3$ & $\phantom{.}10.4$\\
    \midrule
      \multicolumn{13}{c}{LLM pair 2}\\ 
    \midrule
    MT-Bench &  $-0.1$ & $-0.1$ & $-0.1$ & $-0.2$ & $-0.2$ & $-0.2$ & $-0.1$ & $-0.1$ & $\phantom{-}0.0$ & $-0.2$ & $-0.2$ & $-0.2$ \\
    MMLU  &  $\phantom{-}1.6$ & $\phantom{-}4.0$ & $\phantom{-}4.2$ & $\phantom{-}7.9$ & $\phantom{-}5.0$ & $\phantom{-}4.4$ & $\phantom{-}5.0$ & $-2.9$ & $\phantom{-}3.2$ & $\phantom{-}5.2$ & $-0.9$ & $\phantom{-}3.8$ \\
    GSM8K  & $\phantom{.}13.6$ & $\phantom{-}8.7$ & $\phantom{.}18.5$ & $\phantom{.}18.9$ & $\phantom{.}14.4$ & $\phantom{.}18.3$ & $\phantom{.}13.1$ & $\phantom{-}4.0$ & $\phantom{.}15.5$ & $\phantom{.}11.3$ & $\phantom{-}8.4$ & $\phantom{.}10.8$\\
    \midrule
      \multicolumn{13}{c}{LLM pair 3}\\ 
    \midrule
    MT-Bench & $\phantom{-}0.2$ & $\phantom{-}0.0$ & $\phantom{-}0.1$ & $-0.1$ & $-0.1$ & $\phantom{-}0.0$ & $\phantom{-}0.0$ & $\phantom{-}0.2$ & $\phantom{-}0.2$ & $-0.1$ & $\phantom{-}0.1$ & $-0.1$\\
    MMLU & $\phantom{-}5.0$ & $\phantom{-}6.8$ & $\phantom{-}5.8$ & $\phantom{.}11.3$ & $\phantom{-}9.1$ & $\phantom{-}4.7$ & $\phantom{-}8.1$ & $-3.7$ & $\phantom{-}4.8$ & $\phantom{-}7.8$ & $\phantom{-}0.1$ & $\phantom{-}7.2$\\
    GSM8K  & $\phantom{.}20.5$ & $\phantom{.}13.4$ & $\phantom{.}20.9$ & $\phantom{.}24.3$ & $\phantom{.}18.6$ & $\phantom{.}21.6$ & $\phantom{.}17.9$ & $\phantom{.}11.2$ & $\phantom{.}18.9$ & $\phantom{.}16.7$ & $\phantom{.}15.2$ & $\phantom{.}14.2$\\
    \bottomrule
    \end{tabular}
 
\caption{Differences between average benchmark specific scores of responses to
  the original and confounded queries, when the confounder gadget was generated
  for a different surrogate router than the target (black-box setting) for three
  LLM pairs. Positive values indicate a higher average score for responses to
  the confounded queries; higher values are better for the attacker. Results are
  averaged across gadgets. Standard errors were omitted for readability and are
  on average $0.1, 0.8$, and $1.8$ for MT-bench, MMLU and GSM8K, respectively.
  Aligned with the white-box setting, results show almost no decrease in
  performance, and improvement when there is a performance gap for the LLM pair.
} \label{tab:bench_specific_different_router}
\end{table*}

\paragraph{Query-specific gadgets.}
By default, our gadget generation method is query-independent and the same gadget can be used to reroute any query.   An adversary with more resources may instead generate a dedicated gadget for each query (using the same algorithm).

\Cref{tab:transfer_rates_same_router_query_specific} and \Cref{tab:transfer_rates_different_router_query_specific} show the results for the white-box and black-box settings, respectively.
(Here, percentage numbers are not averaged and there is no standard error since we used a single gadget per query.) The white-box results are nearly perfect; the black-box results are often better but sometimes somewhat worse than those for query-independent gadgets. We conjecture that this is due to some level of overfitting.

\begin{table*}[t]
\small
\centering
    \begin{tabular}{l|cccc}
    \toprule 
     & $R_{SW}$ & $R_{MF}$ & $R_{CLS}$ & $R_{LLM}$\\
    \midrule
    MT-Bench & $100$ & $100$  & $100$ & $100$ \\
    MMLU   & $100$ & $\phantom{0}96$ & $100$ & $100$\\
    GSM8K  & $100$ & $100$ & $100$ & $100$\\
    \bottomrule
    \end{tabular}
\caption{Upgrade rates for query-specific gadgets, in the white-box setting. Results are nearly perfect, i.e. nearly all confounded queries are routed to the strong model.} \label{tab:transfer_rates_same_router_query_specific}
\end{table*}

\begin{table*}[t]
\small
\centering
    \begin{tabular}{l|ccc|ccc|ccc|ccc}
    \toprule 
     Surrogate &\multicolumn{3}{c|}{$\hat{R}_{SW}$} & \multicolumn{3}{c|}{$\hat{R}_{MF}$} &\multicolumn{3}{c|}{$\hat{R}_{CLS}$} & \multicolumn{3}{c}{$\hat{R}_{LLM}$} \\
    Target & $R_{MF}$ & $R_{CLS}$ & $R_{LLM}$ & $R_{SW}$ & $R_{CLS}$ & $R_{LLM}$ & $R_{SW}$ & $S_{FM}$ & $R_{LLM}$ & $R_{SW}$ & $R_{MF}$ & $R_{CLS}$\\
    \midrule
    MT-Bench & $100$ & $83$ & $71$ & $100$ & $83$ & $48$ & $100$ & $73$ & $52$ & $100$ & $67$ & $83$\\
    MMLU & $\phantom{0}96$ & $57$ & $89$ & $\phantom{0}95$ & $43$ & $83$ & $\phantom{0}74$ & $13$ & $83$ & $\phantom{0}77$ & $11$ & $30$ \\
    GSM8K & $100$ & $68$ & $74$ & $100$ & $73$ & $68$ & $\phantom{0}81$ & $65$ & $70$ & $\phantom{0}88$ & $54$ & $64$\\
    \bottomrule
    \end{tabular}
\caption{Upgrade rates for query-specific gadgets, in the black-box setting. In most cases results are better than in the query-independent setting, at the cost of a more resource intensive process.} \label{tab:transfer_rates_different_router_query_specific}
\end{table*}

\paragraph{Results for LLM pair 4.} As discussed in~\Cref{sec:exp_setting}, we
replace the strong model that was used by Ong et al.~\cite{ong2024routellm},
GPT-4-1106-preview (rank 28 in the Chatbot Arena leaderboard
\cite{chatbot_arena,chiang2024chatbot}), with the open-sourced Llama-3.1-8B
(rank 58) to reduce the costs of our extensive set of evaluations. In this
section we perform a smaller-scale evaluation of the quality-enhancing attack
performance when using GPT as the strong model, i.e., LLM pair~4. We evaluate
this setting using three of the $n=10$ confounder gadgets for each router.

\Cref{tab:bench_specific_same_router_gpt} shows the results across benchmarks in the white-box setting. Compared to the pair 1 setting (\Cref{tab:bench_specific_same_router}), the attack results in a higher increase in benchmark performance. This further demonstrates higher attack effect on response quality when the performance gap between the weak and strong models is higher.

\begin{table*}[t]
\small
\centering
    \begin{tabular}{l|cc|cc|cc|cc}
    \toprule 
       & \multicolumn{2}{c|}{$R_{SW}$} & \multicolumn{2}{c|}{$R_{MF}$} & \multicolumn{2}{c|}{$R_{CLS}$} & \multicolumn{2}{c}{$R_{LLM}$}\\
     & Original & Confounded & Original & Confounded & Original & Confounded & Original & Confounded\\
    \midrule
    MT-Bench &  $9.2$ & $9.2\pm0.0$ & $9.1$ & $9.3\pm0.0$ & $9.2$ & $9.1\pm0.0$ & $8.9$ & $9.1\pm0.1$ \\
    MMLU & $76$ & $84\pm1\phantom{0}$ & $76$ & $81\pm0\phantom{0}$ & $76$ & $84\pm0\phantom{0}$ & $78$ & $84\pm1\phantom{0}$\\ 
    GSM8K & $62$ & $86\pm0\phantom{0}$ & $65$ & $88\pm1\phantom{0}$ & $68$ & $90\pm2\phantom{0}$ & $66$ & $85\pm2\phantom{0}$\\
    \bottomrule
    \end{tabular}
\caption{Benchmark-specific average scores of responses to the original and confounded queries with GPT-4-1106-preview as the strong model (LLM pair 4), in the white-box setting. Results demonstrate a higher increase in performance with respect to the LLM pair 1 setting, due to the larger performance gap between the models.} \label{tab:bench_specific_same_router_gpt}
\end{table*}

\section{Rerouting Commercial Routers}
\label{sec:commercial}

We evaluate our rerouting attack on several commercial routers: Unify
\cite{unify}, NotDiamond \cite{notdiamond}, OpenRouter \cite{openrouter}, and
Martian \cite{martian}. These routers are available through black-box APIs.
Therefore, we use our black-box attack with the $40$ gadgets optimized for the
open-sourced routers $R_{SW}, R_{MF}, R_{CLS}$, and $R_{LLM}$ ($10$ per router). We perform this
evaluation using the MT-bench benchmark. 

\paragraph{Unify.} 
This router lets users specify a list of models from different providers and a metric configuration for routing decisions.
The available metrics are quality, time to first token, inter-token latency, and
cost. The user can specify the weight for each metric.  Time, latency, and cost
metrics are static and precomputed.  The quality metric is computed for each
query using a neural scoring function that was trained on prompts from several
open datasets (e.g., Open Hermes \cite{OpenHermes}) and labeled using an LLM-as-a-judge \cite{zheng2023judging}.

For our evaluation, we configure the router to choose between GPT-4o
\cite{openai2024gpt4o} as the strong model and Mixtral 8x7B
\cite{jiang2024mixtral} as the weak model.  We focus on the cost and quality
metrics, and set the weight of time and latency to 0 so that they are not
factored into routing decisions. We manually calibrate the weights to $1$ for
the quality metric and $0.02$ for the cost metric.  These weights result in
$49\%$ of the original, unmodified queries being routed to the strong model and $51\%$
to the weak model, resulting in a total cost of $\$0.13$ for the 72
MT-bench queries.  Adding
confounder gadgets generated for the four open-sourced evaluated routers results
in upgrade rates of $79\%, 88\%, 91\%$, and $89\%$, respectively, averaged
across 10 gadgets. The downgrade rate is zero in all cases. In terms of
costs, the addition of the confounder gadget increased the cost to  $\$0.22,
\$0.23, \$0.22$, and $\$0.21$, respectively, averaged across 10 gadgets. 
In other words, the rerouting attack increased the cost of processing the
queries, on average, by a factor of $1.7\times$.

\paragraph{NotDiamond.}
This router lets users route their queries to a list of predefined models.  Available objectives are to maximize quality, or balance quality and cost, or balance quality and latency.  The exact details of the routing logic are not specified.  We focus on cost-aware routing, for which the API docs state that ``NotDiamond will automatically determine when a query is simple enough to use a cheaper model without degrading the quality of the response.''
NotDiamond provides a router selection tool which gives the routing
decision for a particular query without forwarding the query to the chosen
model (thereby incurring no costs). We use this for our evaluation---of course a
real attack would target the NotDiamond API when used for actual routing.

Similar to the Unify experiments, we set GPT-4o as the strong model and
Mixtral-8x7b as the weak model.  Cost-aware routing routes $82\%$ of the
original queries to the strong model, $18\%$ to the weak model. Confounded
queries generated for $R_{SW}, R_{MF}, R_{CLS}$, and $R_{LLM}$ achieve upgrade
rates of $21\%, 18\%, 21\%$, and $15\%$, respectively. The downgrade rates are
$1$--$3\%$.

As opposed to our calibrated routers, NotDiamond aggressively routes to the
stronger model even for unmodified queries in most settings. We tried several
strong/weak model pairs including GPT-4o/Mistral-7B-Instruct-v0.2,
GPT-4o/GPT-4o-mini, and Claude-3-Opus/Claude-3-Sonnet, and observed a similar
20\%--80\% split between strong and weak.

When we changed the strong model to OpenAI's o1-mini and kept Mixtral-8x7b as
the weak model, $54\%$ of the original queries  were routed to the strong model,
$46\%$ to the weak model. In this setting, confounder gadgets yield $13$--$16\%$
upgrade rates and, on average, $3$--$6\%$ downgrade rates.  We conclude that while
the attack is still effective, NotDiamond is more robust than Unify.

\paragraph{OpenRouter.} This framework offers a unified interface for LLMs, and
additionally offers a system that routes users' queries between three specific
models: Llama-3-70b, Claude-3.5-Sonnet, and GPT-4o.  Queries are routed
``depending on their size, subject, and complexity,'' as described in the
documentation.\footnote{\url{https://openrouter.ai/openrouter/auto}}

With OpenRouter, $96\%$ of the original queries are routed to Llama, $4\%$ to
GPT, and none to Claude. Based on the pricing and number of input-output
tokens, the queries' total cost is $\$0.03$ for
processing all evaluated queries. After adding confounder gadgets,
queries originally routed to GPT are still routed to GPT and no queries are ever
routed to Claude.  For queries originally routed to Llama, some gadgets result
in \emph{all} of them being rerouted to GPT, and some have no impact.
Specifically, $4$ out of the $10$ gadgets we optimized using $R_{SW}$ caused all
queries to be rerouted to GPT, $2/10$ using $R_{MF}$, and $3/10$ using
$R_{LLM}$.  None of the gadgets optimized using $R_{CLS}$ had any impact on
routing. In terms of costs, having all queries being rerouted to GPT results with
an average cost of $\$0.25$, a greater than $8\times$ increase over the cost of the original
queries. Given the lack of documentation of the routing algorithm being used,
we are unsure what explains the variability across gadgets.

\paragraph{Martian.} This router is supposed to let the user provide a list of
models and to specify the maximum amount the user is willing to pay for a query
or for 1M tokens. Unfortunately, as of November 14, 2024, the router appears to
ignore the list models provided by the user, and forwards the input to the same
LLM regardless of it. We tested this in settings including one, two, or multiple models.
While responses do not specify which LLM was used, they were identical
across settings, so we excluded Martian from our evaluation.
We notified Martian about the seemingly buggy behavior.

\section{Defenses}

Defenses against rerouting should be cheap.  If the per-query cost of the
defense is comparable to the per-query cost of a strong LLM, deploying the
defense will defeat the main purpose of LLM routing, which is to reduce the cost
of responding to queries.

\paragraph{Perplexity-based filtering.} As explained in
Section~\ref{sec:open-source-results}, perplexity is a measure of how
``natural'' the text looks. Perplexity-based filtering has been suggested in
many contexts as a defense against adversarial text
inputs~\cite{jain2023baseline, alon2023detecting}.  This defense computes the
perplexity of multiple ``trusted'' texts, then compares it with the perplexity
of the suspicious text.  If the latter is significantly higher, or above some
predefined threshold, the text is considered adversarial. Specifically, we
assume the defender has access to a set of unmodified queries. The defender
computes their perplexity values and uses these values to establish a threshold.
Given a new query, the defender checks if its perplexity exceeds the threshold.
If so, the query is flagged as adversarial. The defender can then decide how to
handle such queries. Options include rejecting them or routing them all to the
weak model. Computing the perplexity of a query can be cheap to do, e.g., 
using GPT-2 as we do in this work; this makes it viable for use as a defense 
that doesn't undermine the benefits of routing.

\begin{figure*}[t]
    \centering
    \subfloat[$R_{SW}$]{
    \includegraphics[width=0.24\linewidth]{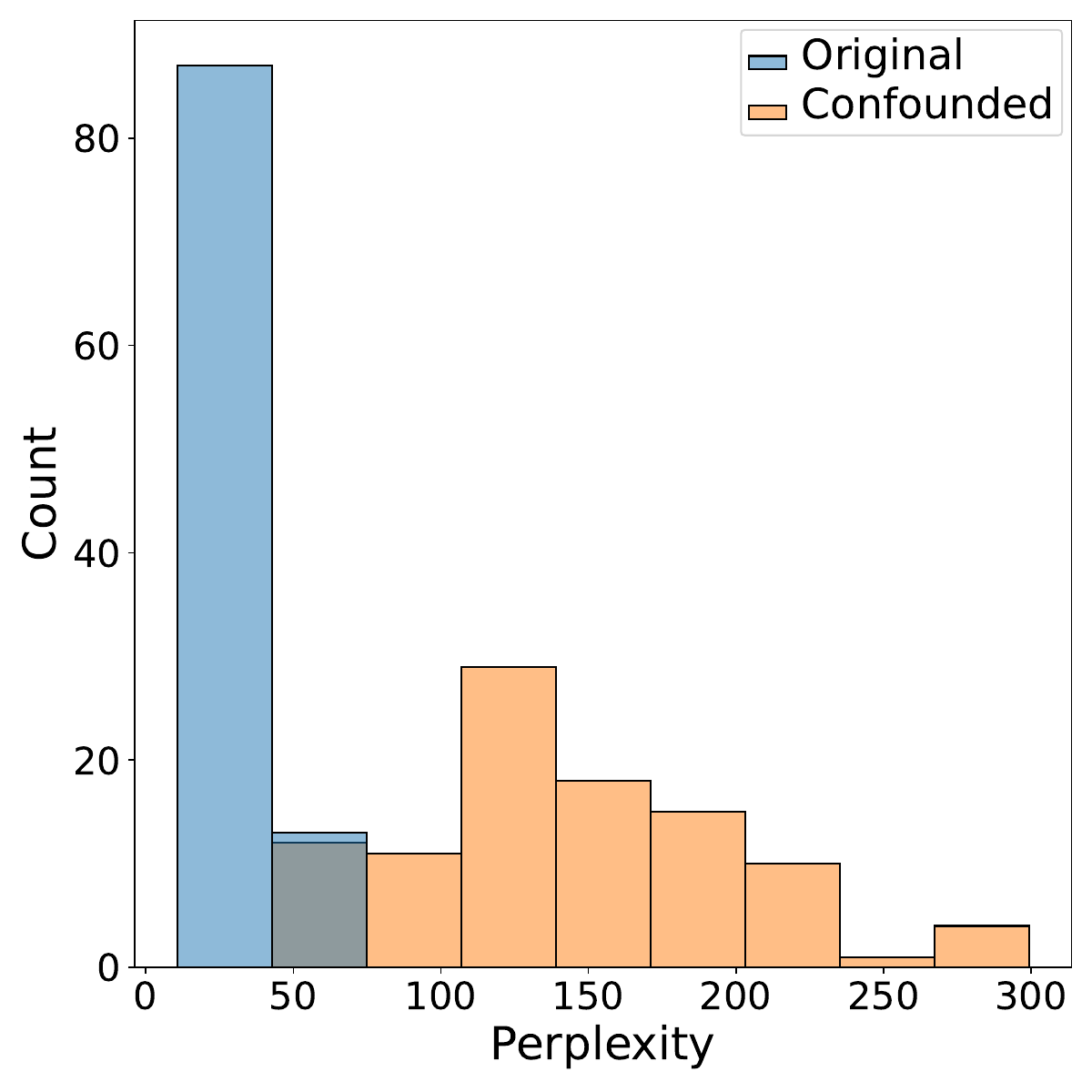}
    }
    \subfloat[$R_{MF}$]{
    \includegraphics[width=0.24\linewidth]{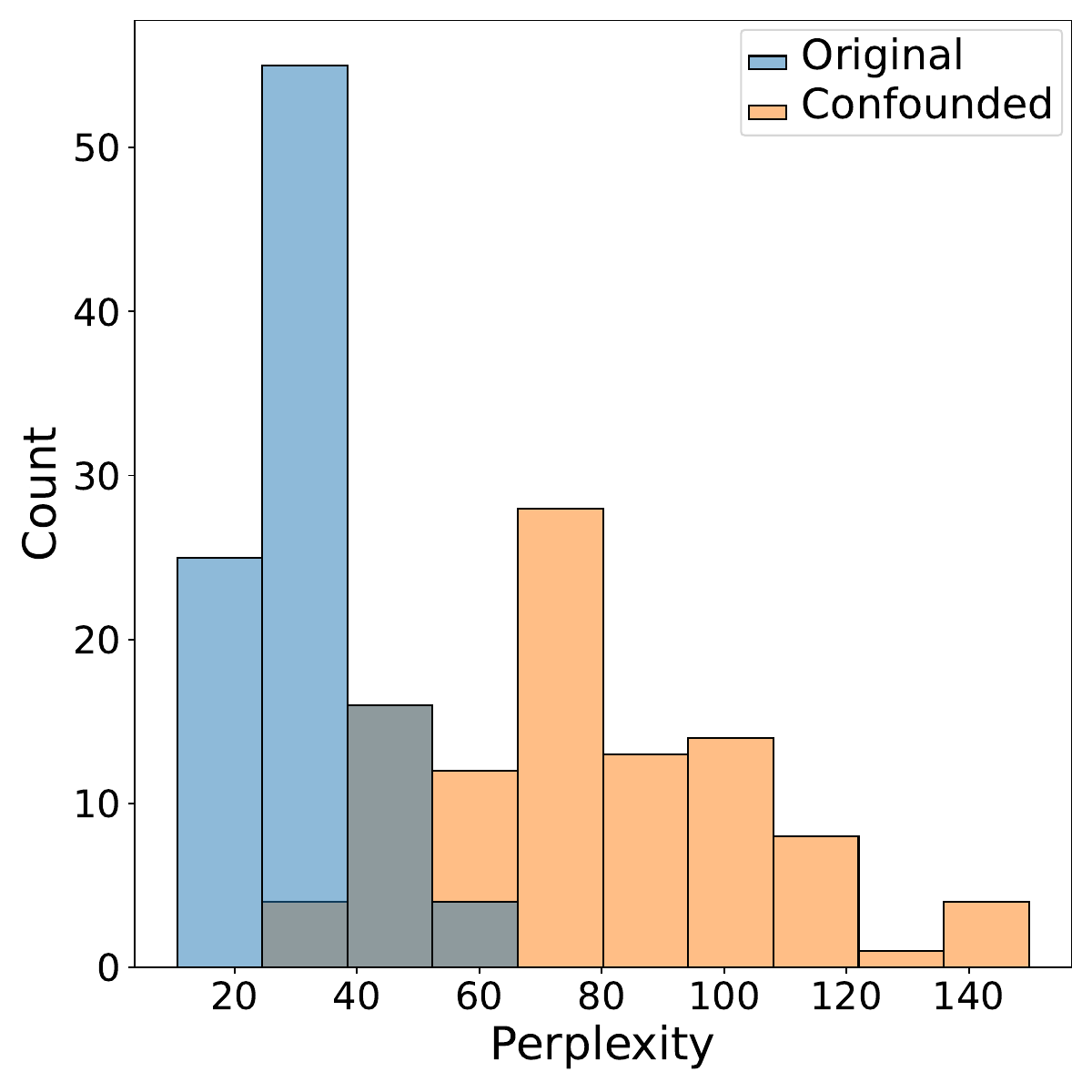}
    }
    \subfloat[$R_{CLS}$]{
    \includegraphics[width=0.24\linewidth]{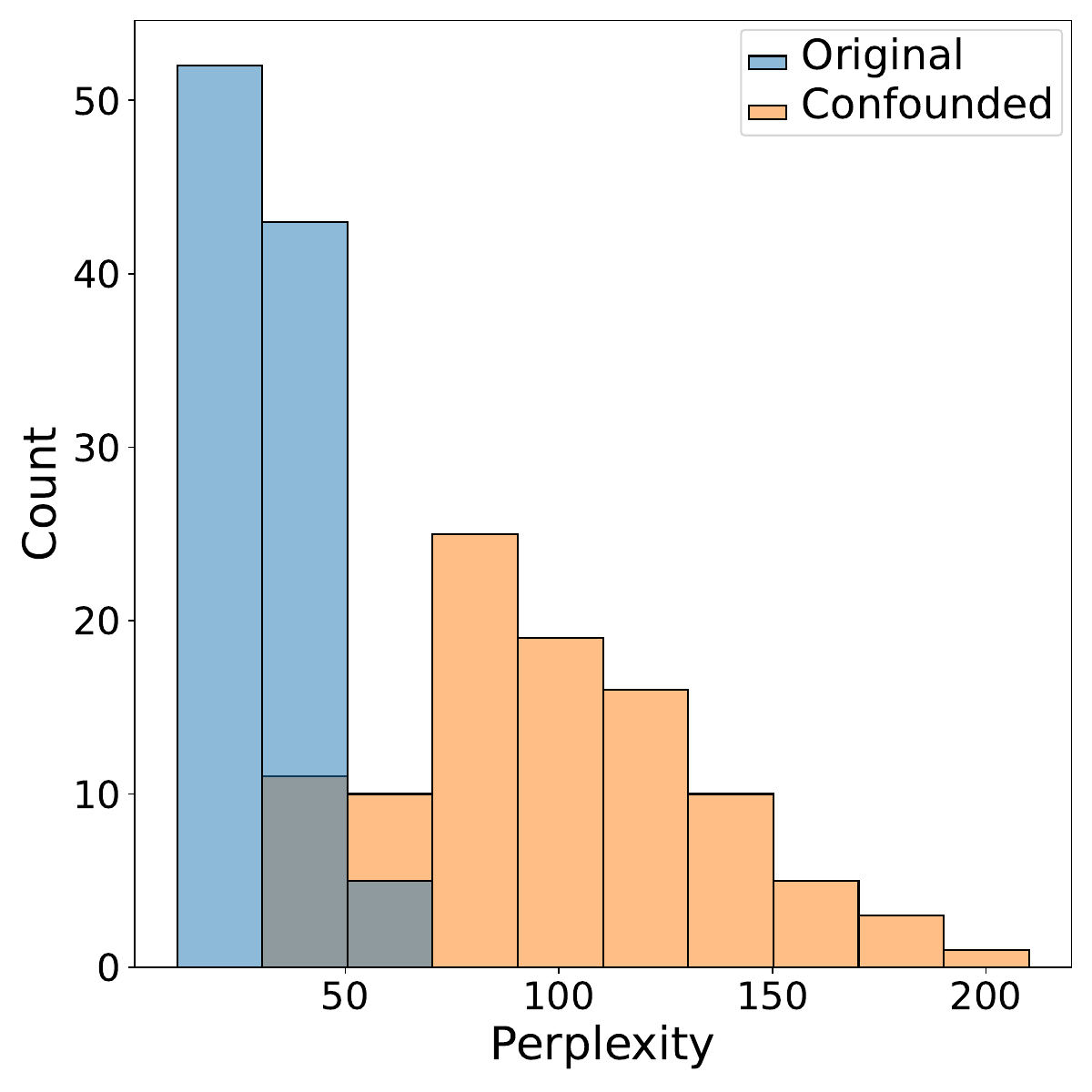}
    }
    \subfloat[$R_{LLM}$]{
    \includegraphics[width=0.24\linewidth]{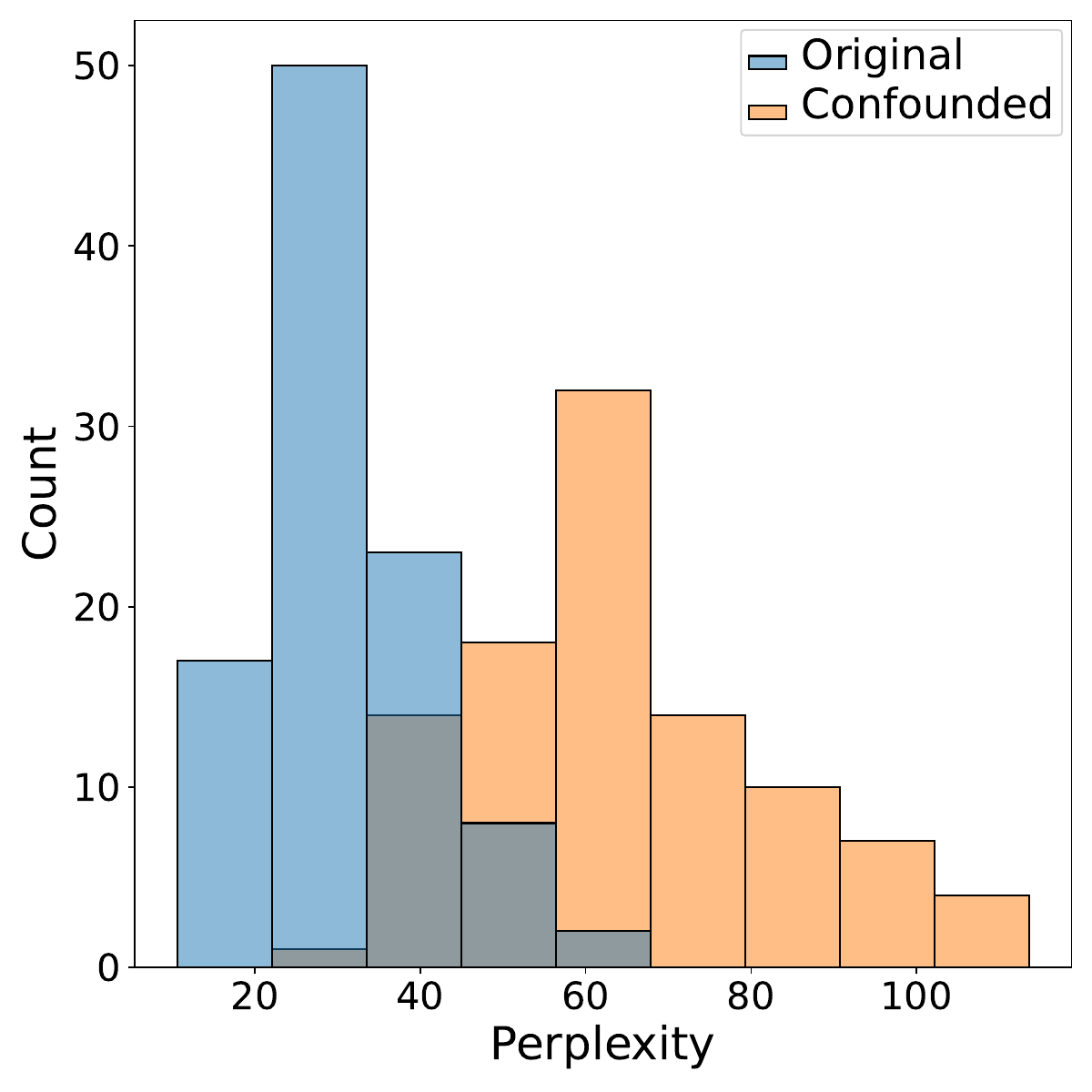}
    }\\
    \subfloat[$R_{SW}$]{
    \includegraphics[width=0.24\linewidth]{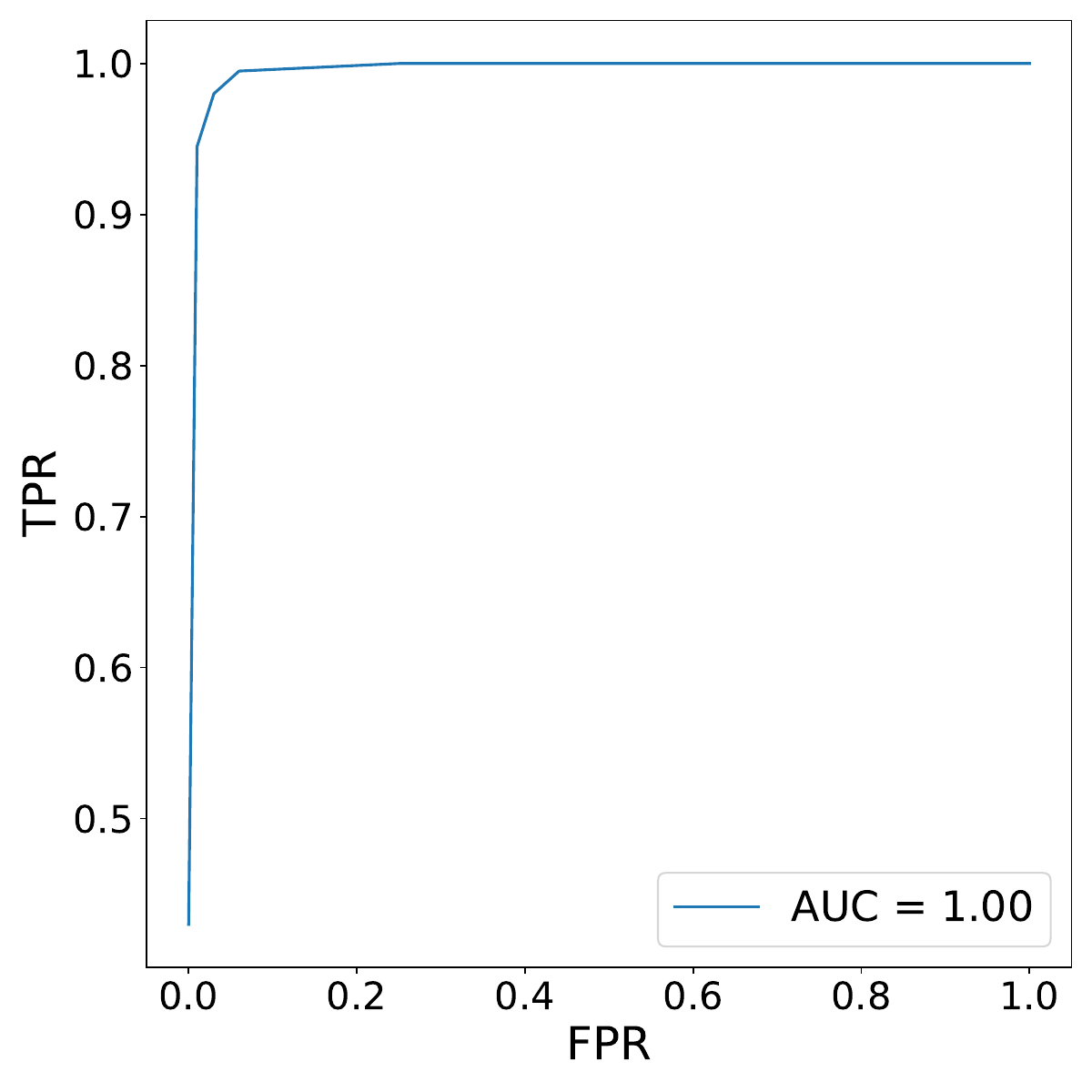}
    }
    \subfloat[$R_{MF}$]{
    \includegraphics[width=0.24\linewidth]{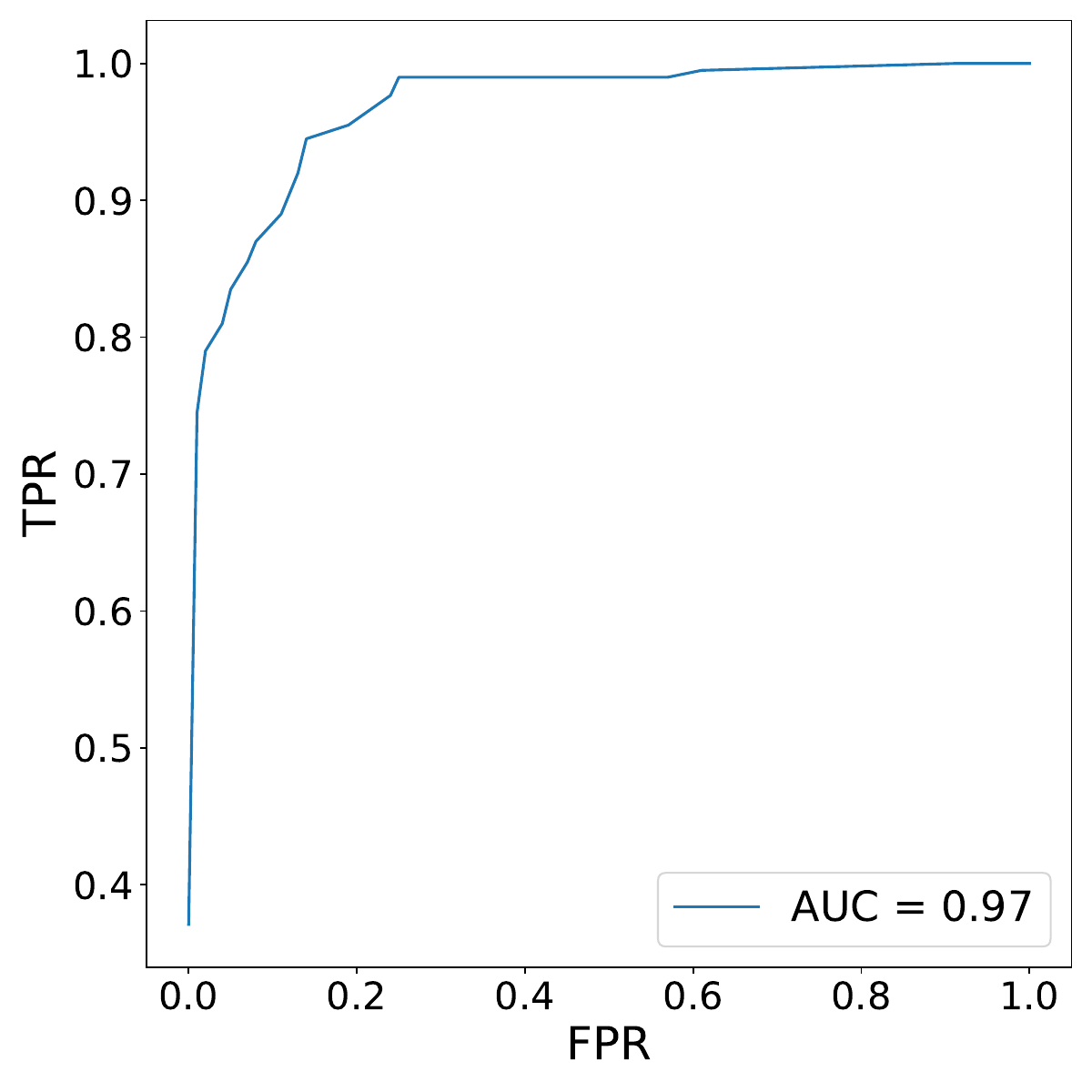}
    }
    \subfloat[$R_{CLS}$]{
    \includegraphics[width=0.24\linewidth]{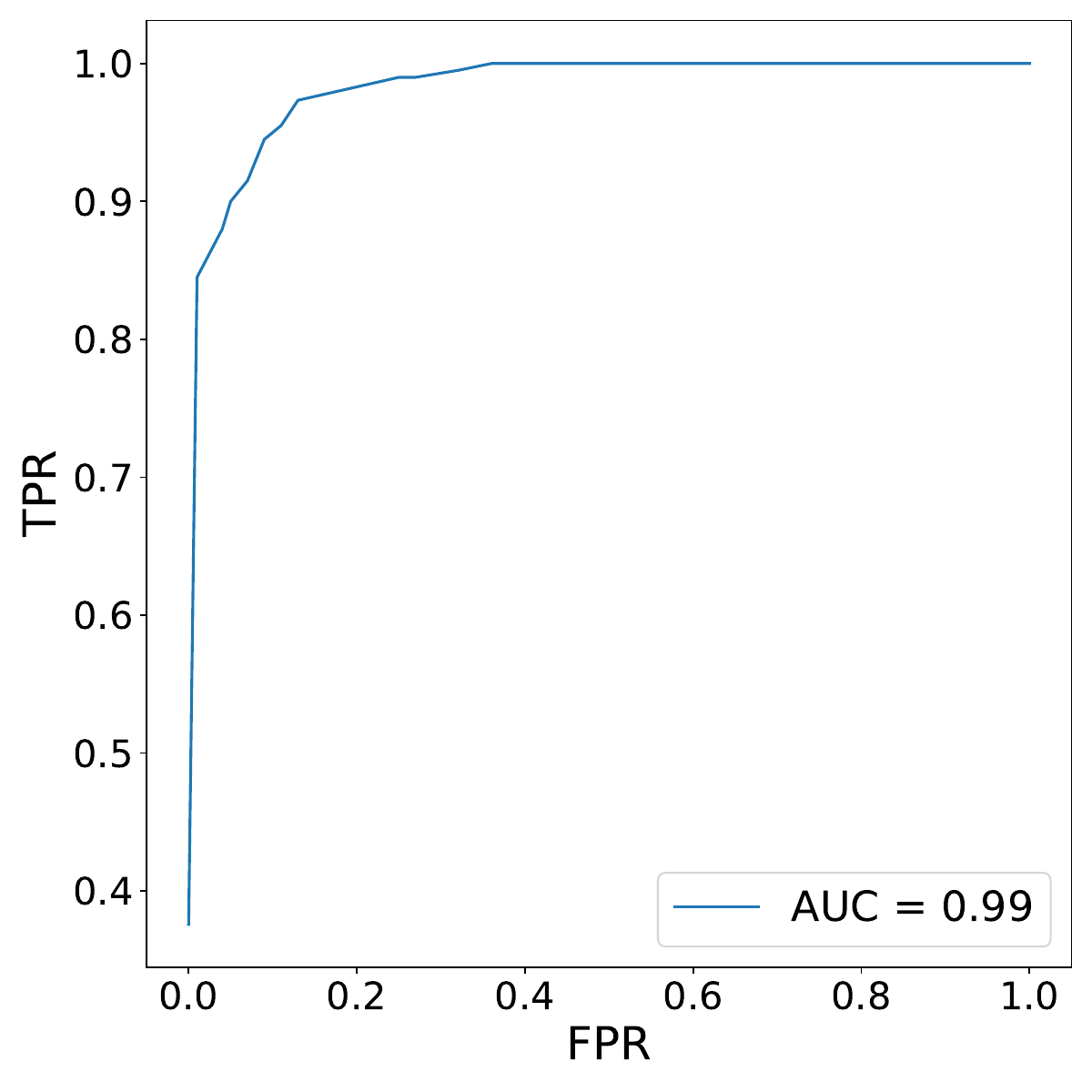}
    }
    \subfloat[$R_{LLM}$]{
    \includegraphics[width=0.24\linewidth]{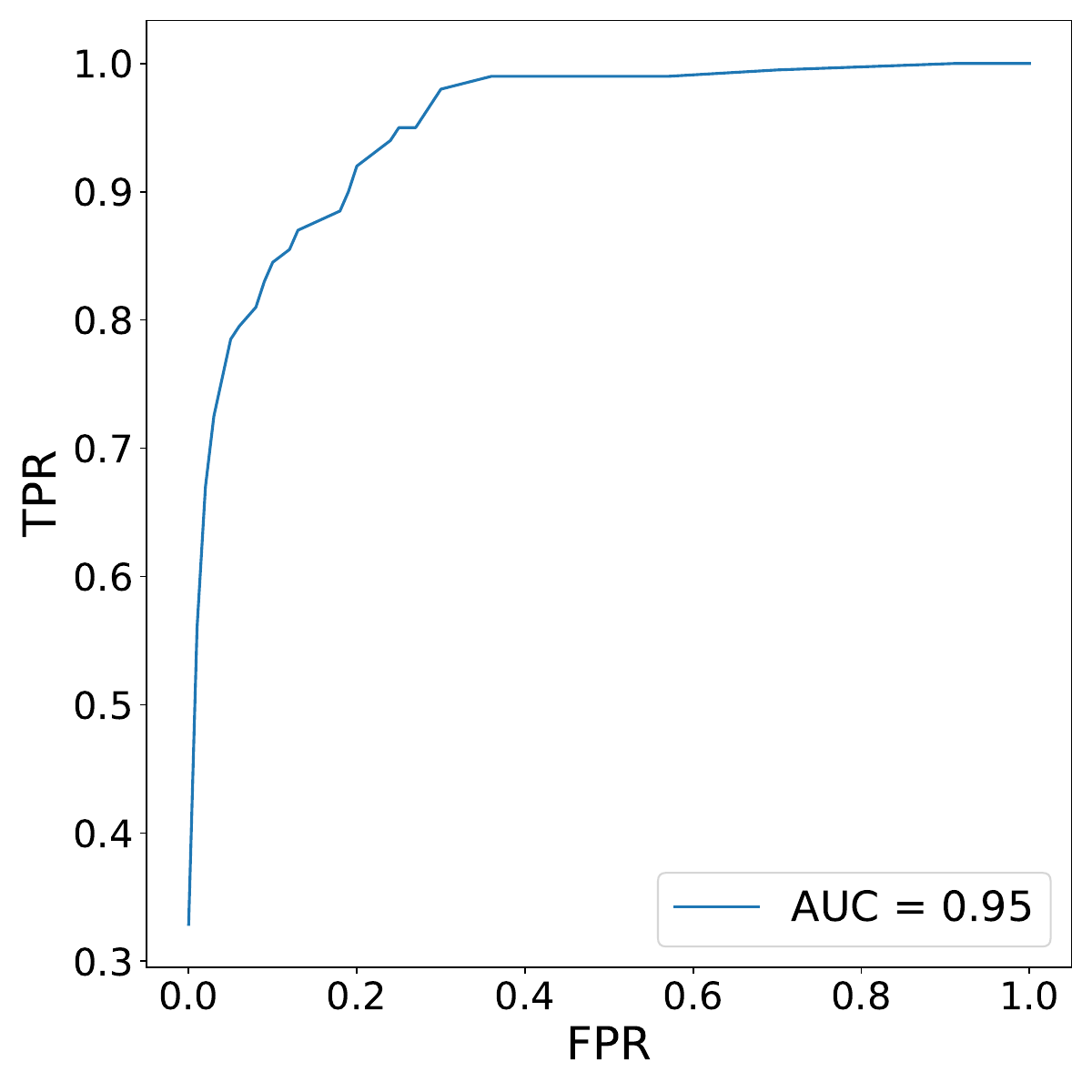}
    }\\

    \caption{Perplexity of the original queries in the GSM8K benchmark compared
    to the perplexity of confounded queries using a single uniformly sampled gadget. We
    additionally present the ROC curve of the defense that detects
    confounded queries by checking if they cross a perplexity threshold, and it's corresponding ROCAUC score.
    Confounded queries have significantly higher perplexity values, and are thus
    easy to recognize and filter out.
    }
    \label{fig:query_ppl_hist_no_opt_gsm8k}
\end{figure*}

To evaluate the effectiveness of such a defense against our attack, we compare the
perplexity values of original and confounded queries.
\Cref{fig:query_ppl_hist_no_opt_gsm8k} presents histograms of perplexity values
for both the original evaluated GSM8K queries and their corresponding confounded
versions, generated using one of the confounder gadgets, sampled uniformly at
random. Additionally, the figure displays the ROC curve for the defense that
detects confounded queries by checking if their perplexity exceeds a threshold.
As can be seen, the confounded queries exhibit significantly higher perplexity
values, making them readily distinguishable from the original queries. For
instance, in the case of the $R_{SW}$ router, setting the threshold value at
$55$ yields a false-positive rate of $3\%$ and a true-positive rate of $97\%$.
Results are similar for other gadgets and benchmarks and were omitted due to
space constraints.

Unfortunately, this defense can be evaded if an adversary incorporates a
perplexity constraint into the gadget generation process. To demonstrate the
feasibility of this evasion strategy, we modify gadget generation to maximize
the score of the routing algorithm $R$ and simultaneously aligning the the gadget's
perplexity to some predefined perplexity value. In more detail, in each iteration $t \in [T]$, we uniformly sample a
target index $j \in [1,n]$ and generate a set $\mathcal{B}$ of $B+1$ candidates
as explained in~\Cref{sec:attack}. We then modify Eq.~\ref{eq:maximize} such
that we now find the candidate that maximizes the difference between the router's score
and the perplexity constraint for the confounder:
\bnm
    \confounder^{(t+1)} \gets \argmax_{\confounder \in \mathcal{B}} \;  \big( S_\theta(\confounder \concat x_i) -\alpha \cdot \lvert\mathsf{PPL}(\confounder)- \PPLcalib\rvert
    \big) \;,
  \enm
where $\mathsf{PPL}(\cdot)$ denotes the perplexity function computed using
GPT-2, the value $\PPLcalib$ denotes a target perplexity value to which we want gadgets'
perplexity to be close, and the value $\alpha$ is a balancing coefficient.  
For the experiments below, we set $\PPLcalib$ to be the
average perplexity value of $100$ uniformly sampled queries\footnote{The perplexity calibration queries were chosen such that they
do not overlap with the queries used for evaluation.} from the GSM8K
benchmark.

\begin{figure*}[t]

    \centering
    \subfloat[$R_{SW}$]{
    \includegraphics[width=0.24\linewidth]{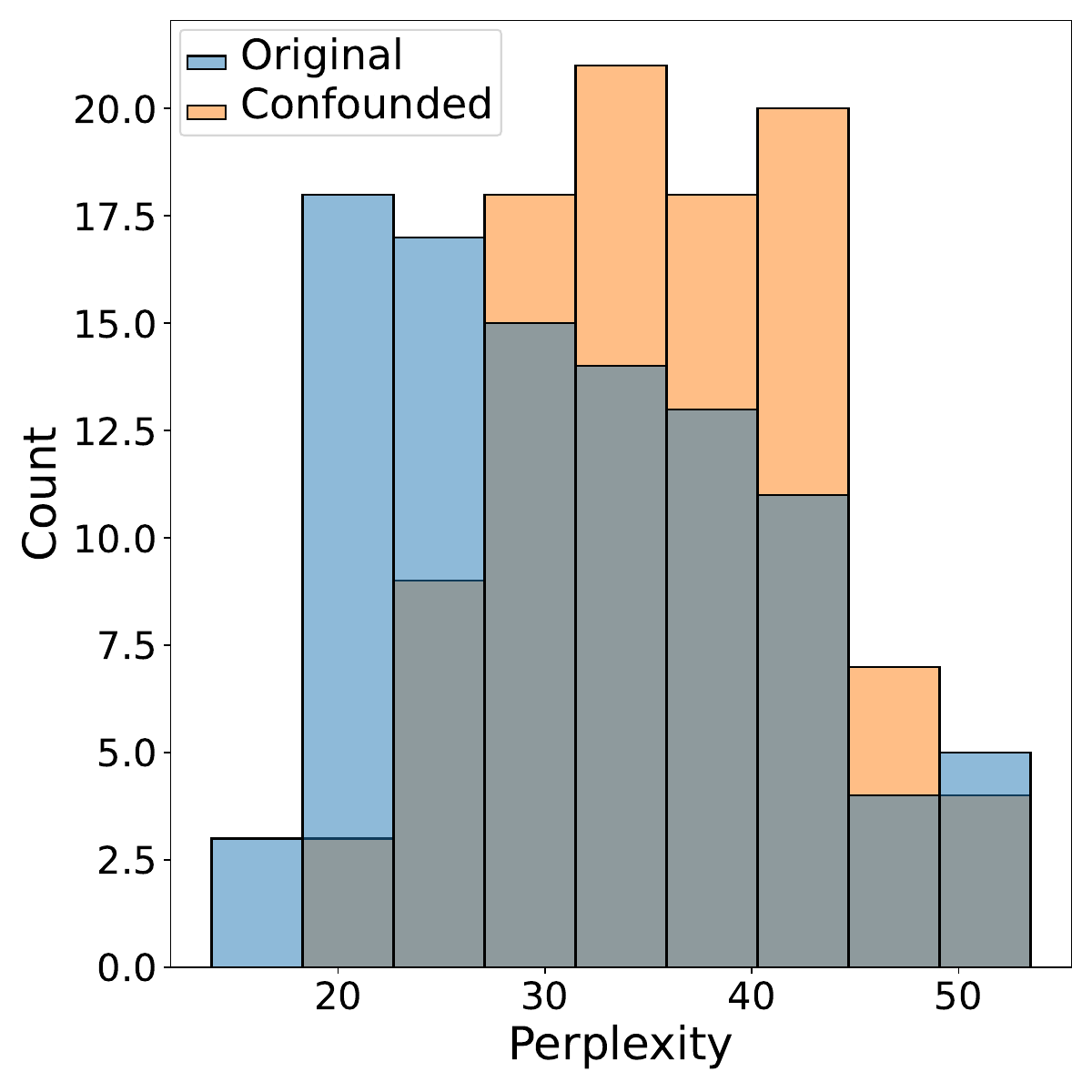}
    }
    \subfloat[$R_{MF}$]{
    \includegraphics[width=0.24\linewidth]{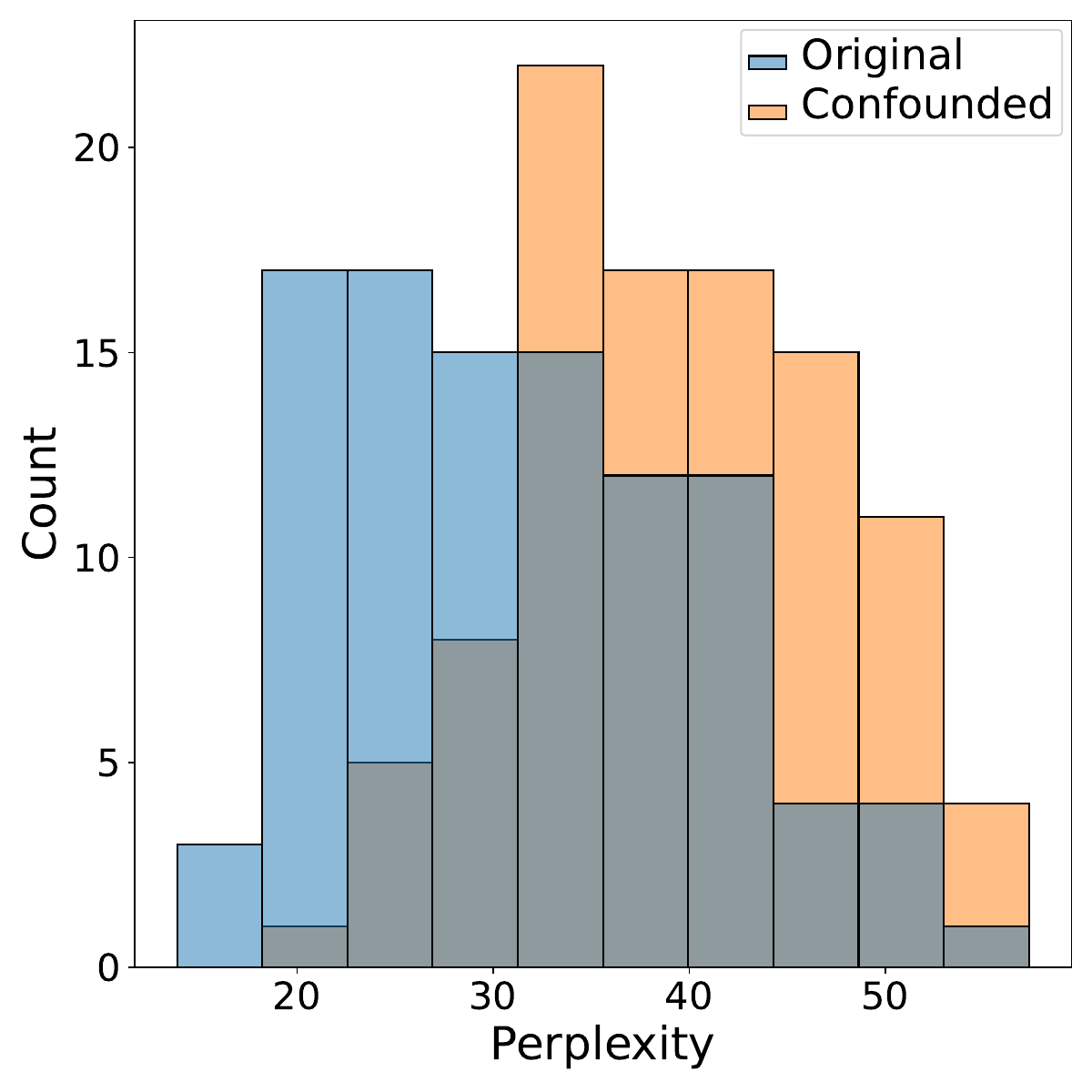}
    }
    \subfloat[$R_{CLS}$]{
    \includegraphics[width=0.24\linewidth]{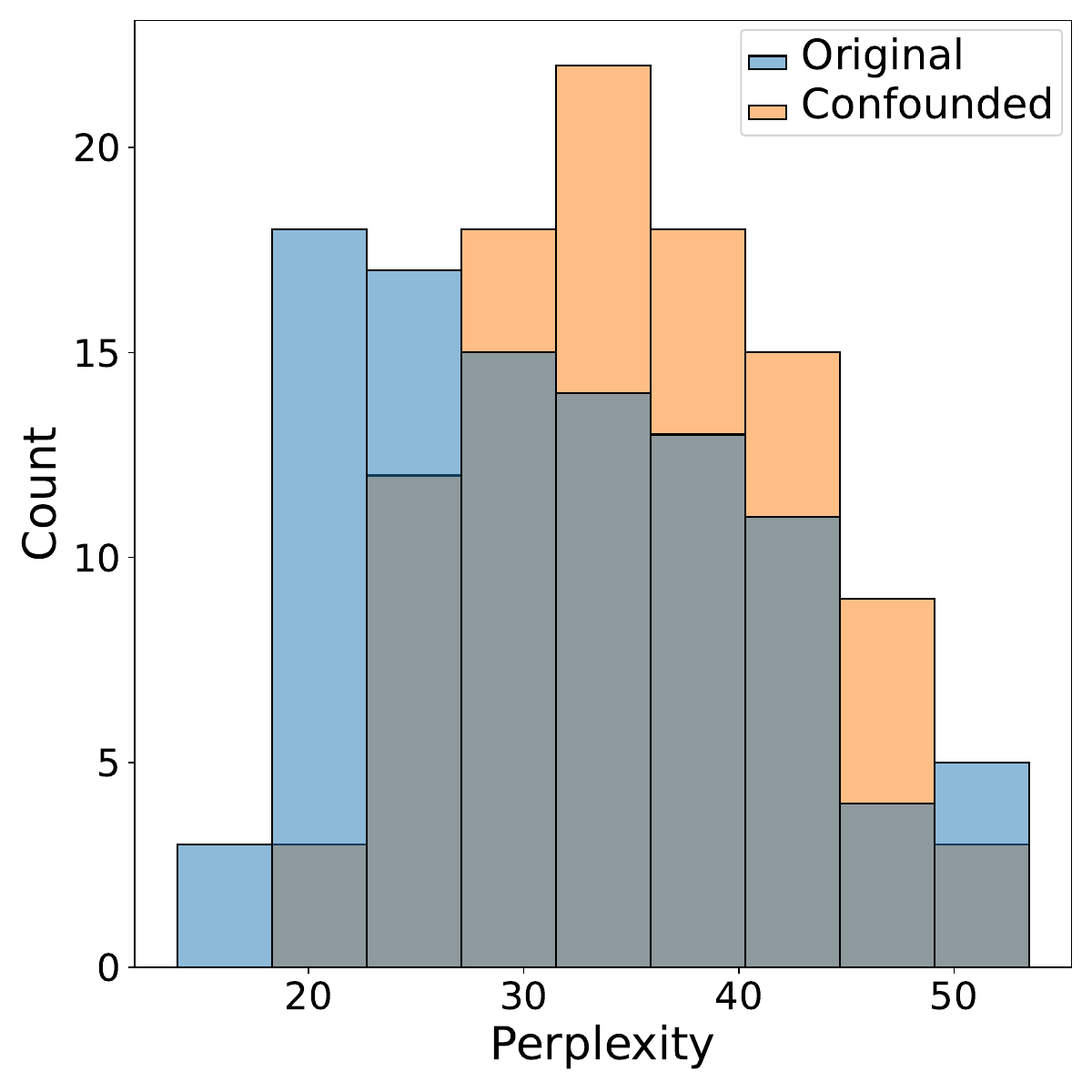}
    }
    \subfloat[$R_{LLM}$]{
    \includegraphics[width=0.24\linewidth]{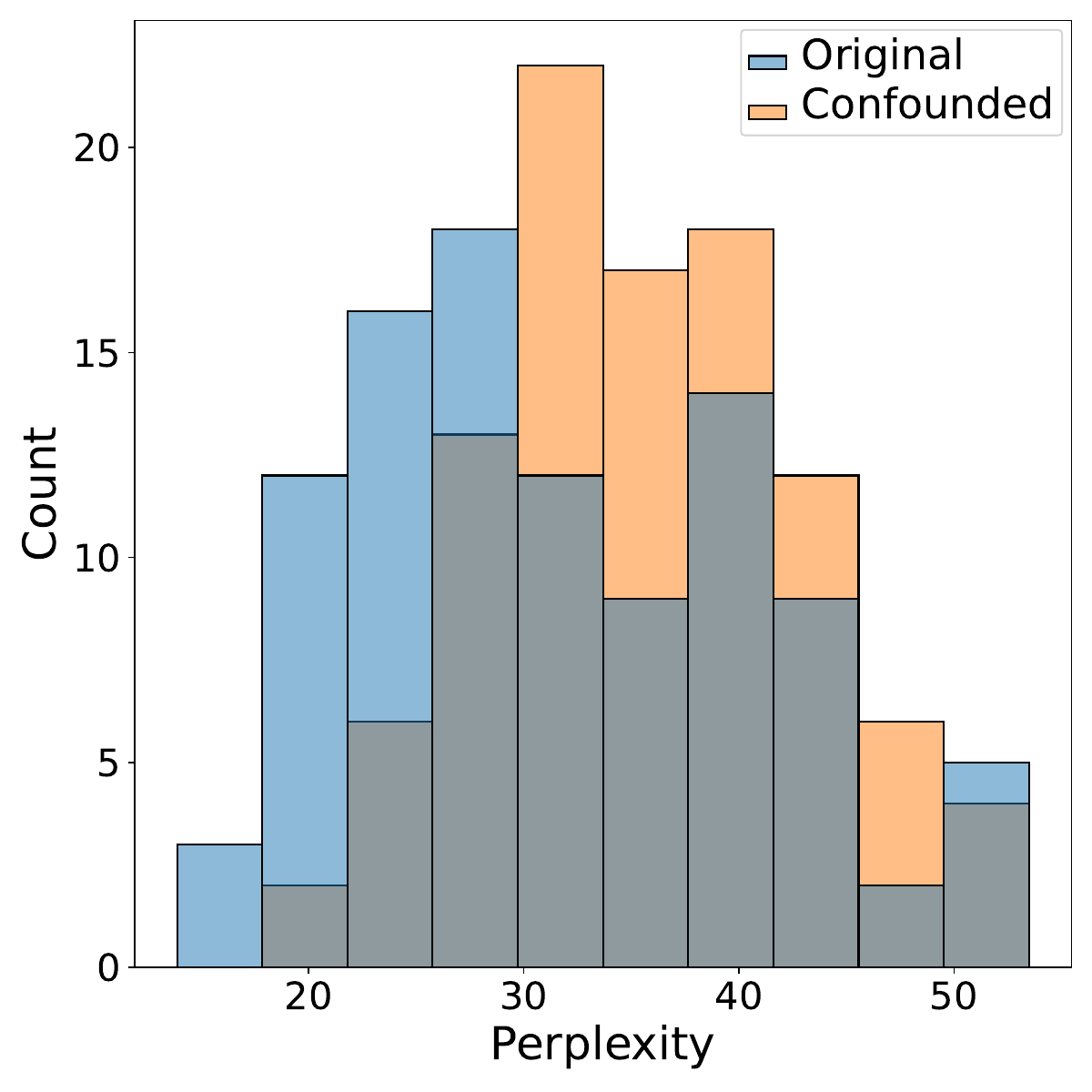}
    }\\
    \subfloat[$R_{SW}$]{
    \includegraphics[width=0.24\linewidth]{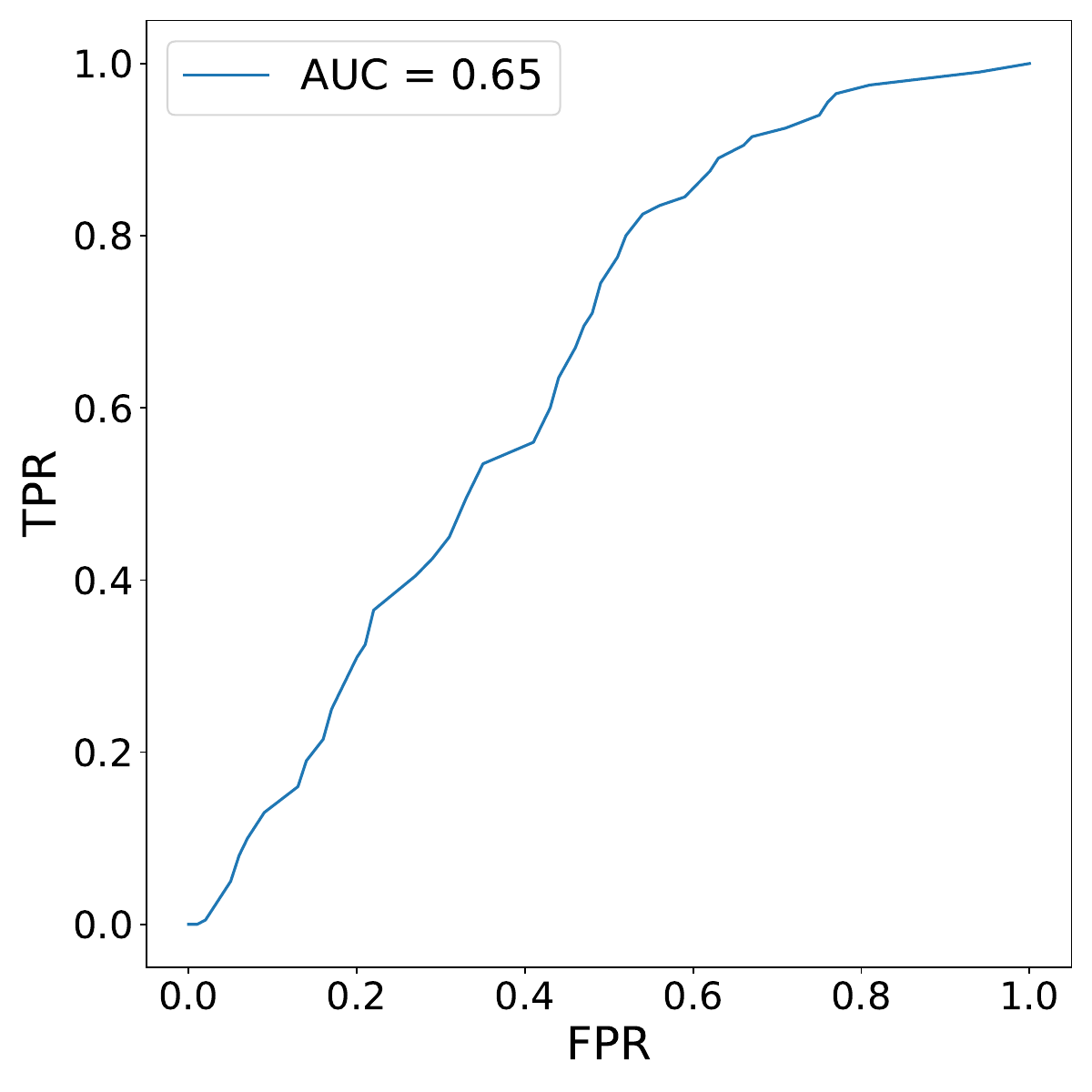}
    }
    \subfloat[$R_{MF}$]{
    \includegraphics[width=0.24\linewidth]{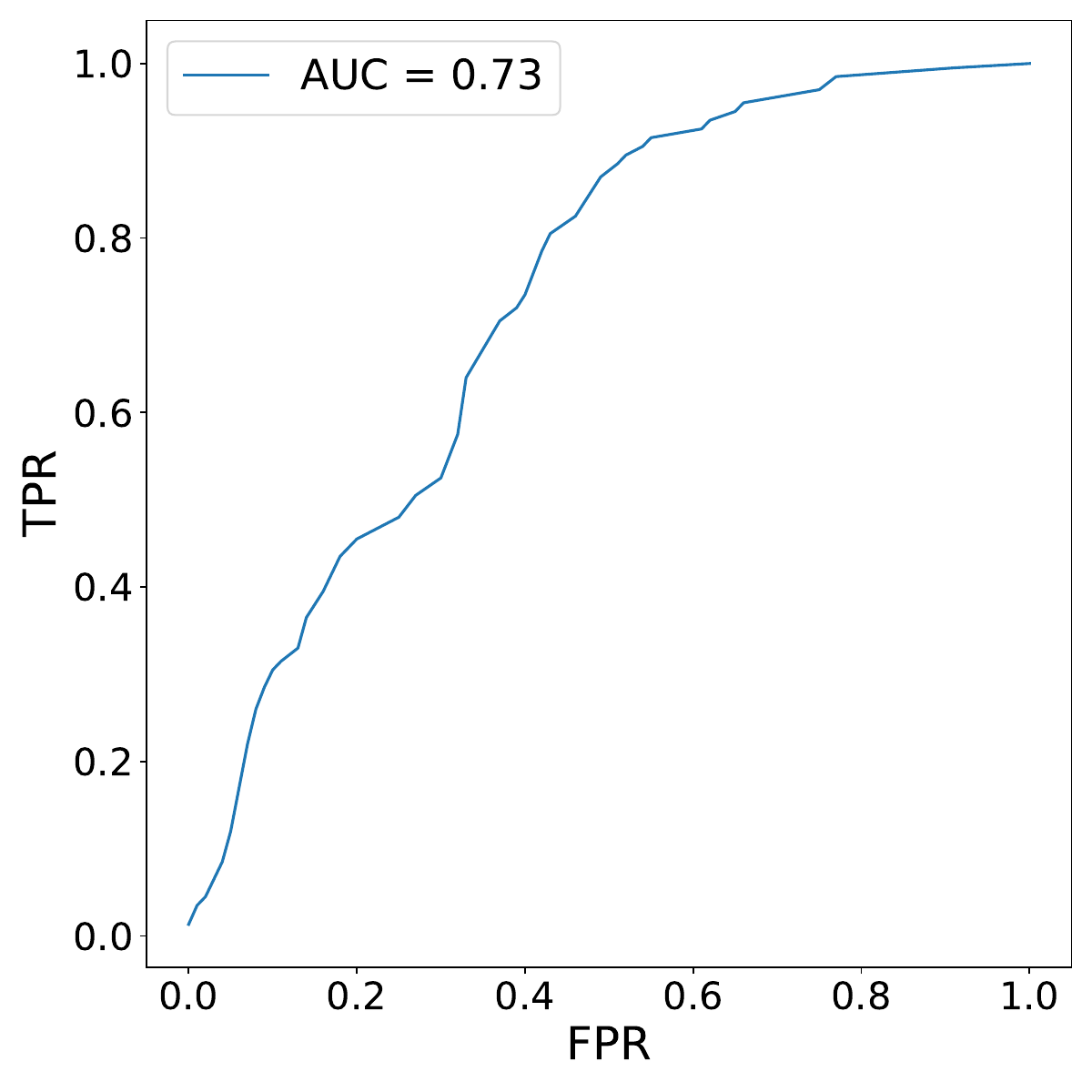}
    }
    \subfloat[$R_{CLS}$]{
    \includegraphics[width=0.24\linewidth]{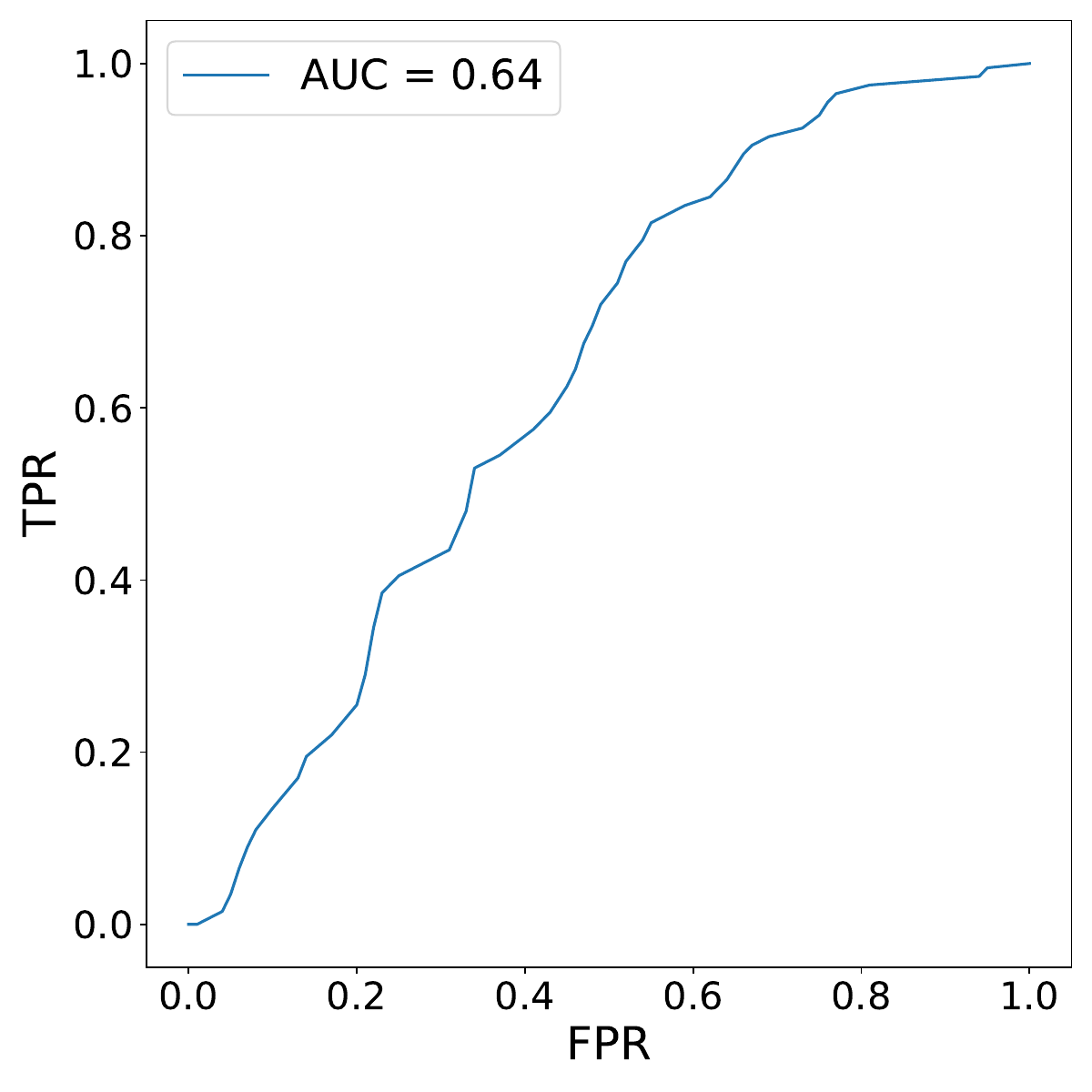}
    }
    \subfloat[$R_{LLM}$]{
    \includegraphics[width=0.24\linewidth]{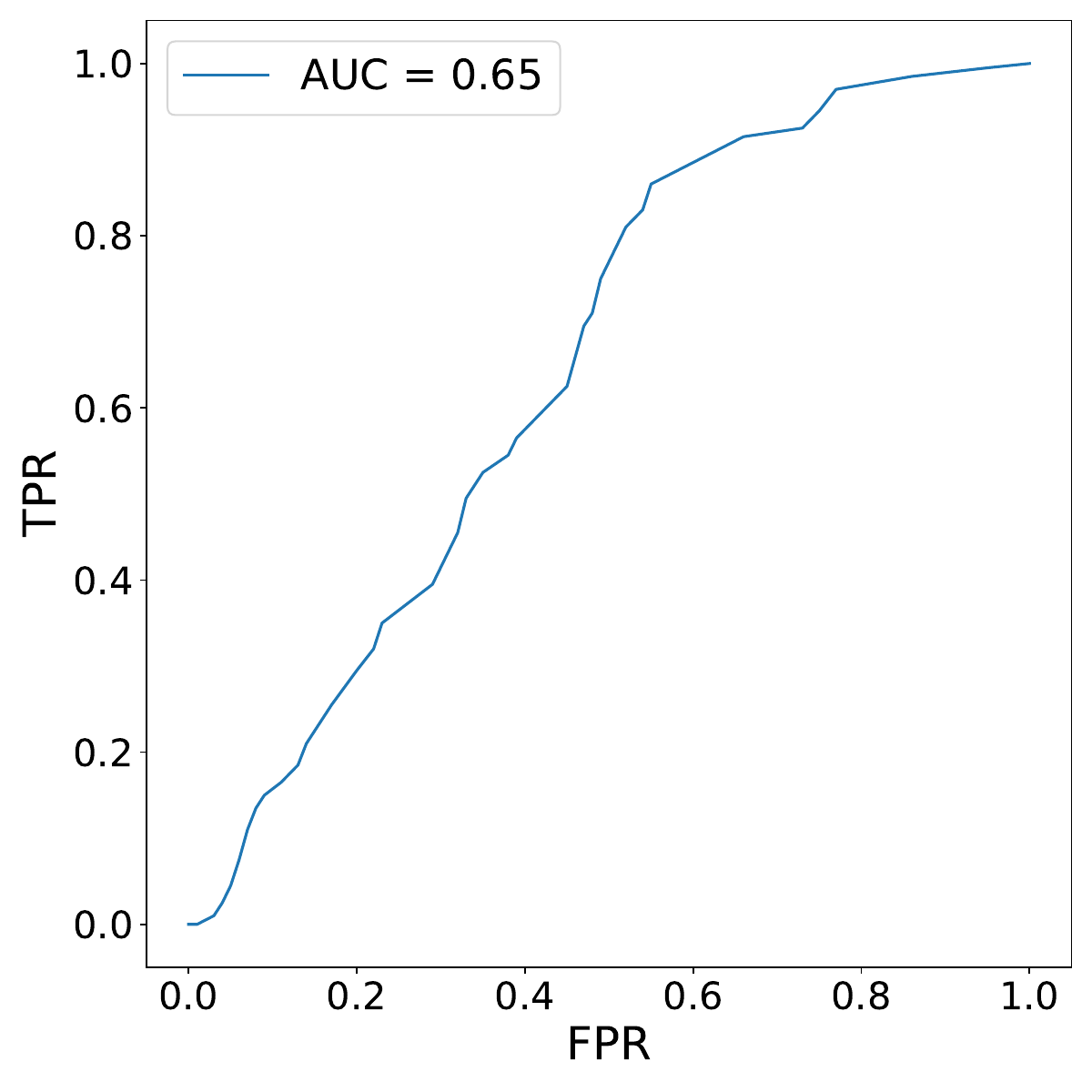}
    }\\

    \caption{Perplexity values of the original and confounded queries, and the corresponding ROC curves of the defense that detects
    confounded queries by checking if they cross a perplexity threshold, when the confounder gadget is optimized for low perplexity, in the GSM8K benchmark and for one gadget sampled uniformly at random. Confounded queries have similar perplexity values as the original queries, and can no longer be easily distinguished based on perplexity alone.}
    \label{fig:query_ppl_hist_with_opt_gsm8k}
\end{figure*}

\Cref{fig:query_ppl_hist_with_opt_gsm8k} shows the results when setting
$\alpha=0.01$, for the GSM8K benchmark and one confounder gadget. The results
demonstrate that modified queries can no longer be easily distinguished from
normal queries by their perplexity alone. For instance, in the case of the
$R_{SW}$ router, setting the threshold value at $55$ as before, no confounded queries
are flagged as anomalous, meaning the true-positive rate is zero.
We note that there is some
variability across gadgets. The average ROCAUC scores of the defense across
ten gadgets with standard deviation indicated  parenthetically, are
$0.66$ ($\pm0.04$), $0.69$ ($\pm0.02$), $0.71$ ($\pm0.02$), and $0.69$
($\pm0.03$) for the $R_{SW},
R_{MF}, R_{CLS}$, and $R_{LLM}$ routers, respectively.  

At the same time, optimizing for low
perplexity does not significantly impact the attack success rate.
\Cref{tab:upgrade_with_ppl_opt} compares the
average upgrade rates (over $n=10$ gadgets) of the original perplexity-agnostic
optimization approach from \Cref{sec:attack} and the perplexity-minimizing one
described above. The attack efficacy might be improvable further by
adjusting~$\alpha$ to find a sweet spot that avoids the defense effectively
while ensuring high rerouting success rate.

The attack is not particularly sensitive to the choice of queries used to obtain the calibration value
$\PPLcalib$. Although $\PPLcalib$ was computed using GSM8K queries, we observe similar
performance when evaluating on the MT-bench and MMLU benchmarks, with average ROCAUC scores of
$0.50$ ($\pm0.01$), $0.51$ ($\pm0.01$), $0.52$ ($\pm0$), and $0.51$ ($\pm0.01$) for MT-bench, and
$0.52$ ($\pm0.03$), $0.54$ ($\pm0.02$), $0.55$ ($\pm0.01$), and $0.53$
($\pm0.02$) for MMLU. One might also try
removing the calibration value altogether, instead simply minimizing the
gadget's perplexity value. However, this can result with an ``overshooting''
effect, where the perplexity value is significantly \emph{lower} than that of
normal queries, thereby making it still distinguishable from standard queries.

In summary, 
perplexity-based filtering is not an effective defense against against
rerouting.

\begin{table*}[t]
\small
\centering
    \begin{tabular}{c|cc|cc|cc|cc}
    \toprule 
       & \multicolumn{2}{c|}{$R_{SW}$} & \multicolumn{2}{c|}{$R_{MF}$} & \multicolumn{2}{c|}{$R_{CLS}$} & \multicolumn{2}{c}{$R_{LLM}$}\\
     & Orig. & PPL-opt. & Orig. & PPL-opt. &Orig. & PPL-opt. & Orig. & PPL-opt. \\
    \midrule
    \midrule
    MT-Bench & $100\pm0$ & $100\pm0$ & $100\pm0$ & $98\pm2$ & $100\pm0$ & $98\pm1$ & $73\pm5$ & $51\pm8$ \\
    MMLU & $\phantom{0}90\pm1$ & $\phantom{0}59\pm5$ & $\phantom{0}78\pm4$ & $74\pm5$ & $100\pm0$ & $\phantom{0}66\pm12$ & $95\pm1$ & $89\pm3$ \\
    GSM8K & $\phantom{0}98\pm0$ & $\phantom{0}70\pm7$ & $100\pm0$ & $98\pm2$ & $100\pm0$ & $88\pm6$ & $94\pm3$ & $81\pm8$ \\
    \bottomrule
    \end{tabular}
\caption{Average upgrade rates for gadgets generated without (``Orig.'') and
  with (``PPL-opt.'') low-perplexity optimization, for the balancing coefficient $\alpha=0.01$. In some cases, optimizing for low perplexity has a negative effect on the attack success rate, however the attack can still be considered successful. A more careful choice of $\alpha$ can potentially limit the effect on the attack success.} \label{tab:upgrade_with_ppl_opt}
\end{table*}

\paragraph{LLM-based filtering.}
Even though adversarially modified queries cannot be easily detected using perplexity,
they may still be ``unnatural.''   A possible defense is to employ an oracle LLM
to determine if the query is natural or not.  This defense requires the router
to invoke an additional LLM for every processed query, which is computationally
expensive in the case of a high-quality open-sourced LLM or financially costly in the case
of a high-quality commercial LLM.  Therefore, this defense is unlikely to be practical.  
Furthermore, it is possible to optimize gadgets so that they both have low perplexity and appear ``natural'' to LLM evaluators~\cite{zhang2024controlled}.

\paragraph{Paraphrasing.} Filtering defenses like those discussed above are passive.  An active alternative
is to paraphrase queries using an oracle LLM.  LLMs are trained to generate
natural text and are thus likely to remove unnatural substrings when
paraphrasing a query.  This defense is likely
impractical for two reasons.  First, and as with LLM-based filtering, it
requires an extra potentially expensive LLM invocation for
each query processed by the router.  Second, it may degrade the quality of responses from the destination LLMs, which are sensitive to the phrasing of queries and prompts.

\paragraph{Detecting anomalous user workloads.} Another possible defense requires the
router to monitor individual user workloads, and identify those users whose queries are routed to the strongest
model with an abnormally high frequency.   The router can
then impose a user-specific threshold.
Of course such workloads may have a benign
explanation, e.g., the user's queries may be unusually complex. Even so, routers
could potentially be designed to perform user-specific routing. For example, one could
imagine using per-user thresholds that are calibrated dynamically to attempt to
maintain a consistent fraction of queries being routed to the strong model. 

Such user-specific routing would complicate implementations, and would make
inaccurate decisions for a user until there is sufficient data about their
queries. The latter is relevant in adversarial settings, since such an approach
would still be circumventable should attackers be able to mount Sybil attacks in
which the attacker creates a new user for, in the limit, each query.

\section{Related Work}

\paragraph{Evasion attacks against ML systems.} A large body of work has investigated
evasion attacks against ML
systems~\cite{dalvi2004adversarial,lowd2005adversarial,szegedy2013intriguing},
also 
referred to as adversarial
examples~\cite{goodfellow2014explaining,papernot2016limitations,papernot2017practical},
and these attacks are now being explored in the context of multi-modal
LLMs~\cite{dong2023robustgooglesbardadversarial} as well as text-only LLMs
(for just one example, see~\cite{cho2024typosbrokeragsback}).
We discussed in \Cref{sec:integrity} how our results compare: LLM control plane
integrity is a distinct AI safety issue, but related in that: (1) control plane
integrity attacks may use evasion-style techniques, 
and (2) control plane integrity attacks might be useful for performing evasion.

\paragraph{Prompt injection against LLMs.} Prompt injection is
a class of attacks against LLMs in which the adversary manipulates the prompt, i.e., the
textual input fed directly to the LLM, causing the LLM to generate outputs that
satisfy some adversarial objective~\cite{perez2022ignore, toyer2023tensor}.
Evasion attacks as discussed above can use prompt injection, 
jailbreaking attacks being a widely explored example in which the adversary aims to bypass some safety
guardrail included in the LLM system, such as ``do not output
expletives''~\cite{liu2023jailbreaking,
schulhoff2023ignore,zou2023universal,wei2024jailbroken,
zhu2023autodan,chu2024comprehensive}.

Prompt
injection is also used for extraction attacks that aim to infer some information from or
about the model, for example, the system
prompt~\cite{perez2022ignore,zhang2023prompts,schulhoff2023ignore}, training
data samples \cite{nasr2023scalable}, or model
parameters~\cite{carlini2024stealing}. In indirect prompt injection attacks~\cite{greshake2023not}, the adversaries do
not directly interact with the target LLM, and instead inject adversarial inputs
into third-party data, which is then added to the LLM prompt (intentionally or
unintentionally) by the victim application and/or its users. This relates to
another category of attacks that target LLM-based applications, such as RAG
systems, and invalidate their integrity by exploiting the weaknesses of the
underlying LLM \cite{chaudhari2024phantom,shafran2024machine}.

Our attacks also modify queries, but with a different aim than the above
types of attacks:
undermining the integrity of the control plane routing, rather than the LLM
itself. Future work might
investigate indirect control plane integrity attacks that, analogously to
indirect prompt injection, serve to somehow trick users of a 
routing system into forming control-plane-confounding queries.

\paragraph{Attacks against MoE.} Mixture-of-Experts (MoE) architectures enable
using multiple expert modules for processing a given query with a lower
computational cost by including an inner routing mechanism that in every layer
routes different tokens to a small number of experts
\cite{du2022glam,fedus2022switch,riquelme2021scaling,shazeer2017outrageously}.
This can be thought of as an internal router within a single LLM, rather than 
an external control plane that orchestrates multiple LLMs. 
MoE has increased in popularity as it allows to build larger models at a fixed
compute budget---not all parameters are used at the same time. 

Hayes et al.~\cite{hayes2024buffer} identified a 
vulnerability in MoE that can be exploited for a denial-of-service attack against
MoE. Thus control plane integrity issues appear to extend to the context of single-LLM
MoE systems, and future work could explore this connection further.

Yona et al.~\cite{yona2024stealing} presented a side-channel attack on MoE
that enables an attacker to reveal other users' prompts. We expect that
side-channel attacks against LLM control planes exist as well, for example, to
infer which models are used via timing of responses. Such attacks, which
target confidentiality, are outside the scope of control plane integrity.

\section{Conclusion}

LLM routers balance quality and cost of LLM inference by routing different
queries to different LLMs.  They are an example of a broader, emerging class of
systems we call ``LLM control planes'' that aim to achieve various quality,
efficiency, and cost objectives by orchestrating use of multiple LLMs to respond
to a query.

We introduced and defined a new safety property, \emph{LLM control plane
integrity}.  Informally, this property holds if an adversarial user cannot
influence routing decisions made by the control plane.  To show that existing
LLM routers do not satisfy this property, we designed, implemented, and
evaluated a black-box optimization method for generating query-independent
``confounder gadgets.''  When added to any query, the confounder gadget confuses
the router into routing the query to the adversary-chosen LLM.

We evaluated the efficacy of confounder gadgets on multiple open-source and
commercial routers and demonstrated that they successfully reroute queries
without a negative impact on the quality of responses.  We also discussed
defenses against these attacks and indicated directions for future research.

\section*{Acknowledgments}

This research was supported in part by the Google Cyber NYC Institutional Research Program, the Israel Science Foundation (Grant No.\ 1336/22), and the European Union (ERC, FTRC, 101043243). Views and opinions expressed are however those of the author(s) only and do not necessarily reflect those of the European Union or the European Research Council. Neither the European Union nor the granting authority can be held responsible for them.

\clearpage

\bibliographystyle{IEEEtranS}
\bibliography{bibliography}

\appendix
\section{Gadget Examples}
\label{sec:substring_examples}
Below are a few examples of the optimized confounder gadgets $\confounder$ we got, for the different routing algorithms:
\begin{itemize}
    \item $R_{SW}$:
    \promptbox{legationbelongs967reglo’hui\\(DictionaryizedNameantal bidi.numberOf}
    \promptbox{demanding\_RESULTS! XVI-ioibili Written Hormchriftlatable}
    \item $R_{MF}$:
    \promptbox{041 personne.nama tomeTN Exact relação tinha seis.Description}
    \promptbox{NUMAGEMENTMedian157\_mut-camerauacao nije Hir Ter}
    \item $R_{CLS}$:
    \promptbox{contiene DEM USERNAME fourteen chees781 consultants200 inici DOJ}
    \promptbox{571:\\Ord:nth Norwegian Mercer\_docs Abr226\_METADATA}
    \item $R_{LLM}$:
    \promptbox{dated:frameifyumi345 Kurdasciiuzeiphertext}
    \promptbox{Midnightexecution431!784 below1 unwrap  : /\ n /\ n}
\end{itemize}

\section{Ablation Study}
\label{sec:ablations}

In this section,
we evaluate the effect of different hyperparameters and design choices (in the white-box setting). 

\paragraph{Prefix vs.\ suffix.}
\label{sec:suffix}
As described in~\Cref{sec:attack}, we prepend the confounder gadget to the query.  An alternative is to append it.  This is straightforward for MT-bench and GSM8K, but MMLU consists of multi-choice questions followed by a list of possible answers, and the term ``Answer:''. We insert the gadget at the end of the question text and before the possible answers.  If we append it at the very end, after ``Answer:'', the LLM assumes the query was answered and in many cases does not generate any output at all. 

\Cref{tab:concat_method_ablation} shows that average upgrade rates are similar regardless of whether the gadget was inserted as a prefix or a suffix. For MMLU, prefix works better.
The downgrade rate is $0\%$ in all cases.

\begin{table*}[t]
\small
\centering
    \begin{tabular}{l|c|cccc}
    \toprule 
     & & $R_{SW}$ & $R_{MF}$ & $R_{CLS}$ & $R_{LLM}$ \\
    \midrule
    \midrule
    \multirow{2}{*}{MT-Bench} & Prefix & $100\pm0$ & $100\pm0$ &  $100\pm0$ & $\phantom{0}73\pm5$ \\
    & Suffix  & $100\pm0$ & $100\pm0$ & $100\pm0$ & $\phantom{0}84\pm4$ \\
    \midrule
    \multirow{2}{*}{MMLU} & Prefix & $\phantom{0}90\pm1$ & $\phantom{0}78\pm4$ & $100\pm0$ & $\phantom{0}95\pm1$ \\
    & Suffix  & $\phantom{0}82\pm2$ & $\phantom{0}63\pm3$ & $\phantom{0}93\pm1$ & $\phantom{0}93\pm1$\\
    \midrule
    \multirow{2}{*}{GSM8K} & Prefix & $\phantom{0}98\pm0$ & $100\pm0$ & $100\pm0$ & $100\pm0$ \\
    & Suffix  & $\phantom{0}94\pm1$ & $100\pm0$ & $100\pm0$ & $94\pm3$ \\
    \bottomrule
    \end{tabular}
\caption{Average upgrade rates for different ways of adding the gadget to queries, in the white-box setting. Results are similar in both methods, with a slight preference to the prefix approach.} \label{tab:concat_method_ablation}
\end{table*}

As mentioned in~\Cref{sec:exp_setting}, to encourage the LLMs to follow the specific format in their responses (so they can be parsed and compared with the ground-truth answers), we add a short prefix to the MMLU and GSM8K queries that instructs the model how to respond.  We phrase this instruction as follows: ``\emph{Answer the question using the format: ``Answer: [A/B/C/D]. Explanation: [EXPLANATION]''}'' for the multi-choice queries of the MMLU benchmark, and a similar version for GSM8K. We add this instruction after modifying the queries with the confounder gadget, i.e. the instruction is prepended to the gadget.

An alternative to insert the instruction after the gadget but before the query, however we observed this to slighly underperform its counterpart.
In the white-box setting we observe a slight decrease in the average (across all four routers) upgrade rate from $91\%$ to $89\%$ for the MMLU benchmark, and from $98\%$ to $91\%$ for the GSM8K benchmark.
In the black-box setting, the average upgrade rate on MMLU reduces from $57\%$ to $49\%$ and on GSM8K from $73\%$ to $64\%$. 

\paragraph{Token sampling method.}
When generating the confounder gadget (see~\Cref{sec:attack}), we iteratively replace tokens with the goal of maximizing the routing algorithm's score for the gadget.  Candidate replacement tokens are chosen uniformly at random. An alternative is to choose candidates based on their probability of appearing in natural text.  To evaluate this method, we compute token probabilities by parsing and tokenizing the wikitext-103-raw-v1 dataset~\cite{merity2016pointer}. 

\Cref{tab:token_sampling_method_ablation} shows that in most cases uniform sampling of replacement tokens yields better upgrade rates.  We conjecture that uniform sampling produces more unnatural text, confusing the router.  For example, for the $R_{SW}$ routing algorithm, uniform sampling produces the following gadget: ``\emph{legationbelongs967reglo’hui(DictionaryizedNameantal bidi.numberOf}'', whereas sampling according to natural probabilities produces ``\emph{total occurred According number Letar final Bab named remainder}''. 

\begin{table*}[t]
\small
\centering
    \begin{tabular}{l|c|cccc}
    \toprule 
     & & $R_{SW}$ & $R_{MF}$ & $R_{CLS}$ & $R_{LLM}$ \\
    \midrule
    \midrule
    \multirow{2}{*}{MT-Bench} & Uniform &  $100\pm0$ & $100\pm0$ & $100\pm0$ & $\phantom{0}73\pm5$ \\
    & Natural Prob.  & $100\pm0$ & $\phantom{0}97\pm2$ & $100\pm0$ & $\phantom{0}70\pm5$\\
    \midrule
    \multirow{2}{*}{MMLU} & Uniform &  $\phantom{0}90\pm1$ & $\phantom{0}78\pm4$ & $100\pm0$ & $\phantom{0}95\pm1$  \\
    & Natural Prob. &  $\phantom{0}77\pm2$ & $\phantom{0}41\pm3$ & $\phantom{0}96\pm2$ & $\phantom{0}87\pm4$ \\
    \midrule
    \multirow{2}{*}{GSM8K} & Uniform & $\phantom{0}98\pm0$ & $100\pm0$ & $100\pm0$ & $\phantom{0}94\pm3$  \\
    & Natural Prob. & $\phantom{0}88\pm2$ & $\phantom{0}92\pm3$ & $100\pm0$ & $\phantom{0}83\pm9$ \\
    \bottomrule
    \end{tabular}
\caption{Average upgrade rates for different ways of sampling candidate tokens during gadget generation, in the white-box setting. Uniformly sampling the tokens yields better upgrade rates in most cases.} \label{tab:token_sampling_method_ablation}
\end{table*}

\paragraph{Number of tokens in the gadget.}
In our main evaluation, the gadgets are composed of $n=10$ tokens. We evaluate the effect of using less ($n=5$) or more ($n=20$ or $n=50$) tokens. We observed that 5 tokens were insufficient to make changes to the routing algorithm's score and thus we were not able to optimize the gadget in this setting. As for 20 tokens, we observe a a small improvement in the white-box setting, increase the average upgrade rate from $93.9\%$ to $95.8\%$, and a bigger improvement in the black-box setting, increase the average upgrade rate from $70.2\%$ to $81.3\%$. Using 50 tokens further increases the upgrade rates, to $98.2\%$ in the white-box setting and $84.2\%$ in the black box setting. The average convergence rate increases as well, from $60$ iterations for $10$ tokens, to $70$ for $20$ tokens, and $100$ for $50$ tokens. Overall this evaluation suggests that our rerouting attack can be even further improved by using longer gadgets, however it is important to be careful not to make them too long to the point that they might degrade the performance of the underlying LLM.

\section{Optimization-Free Gadget Generation}
\label{sec:other_methods}

We evaluate optimization-free alternatives to our black-box optimization method for generating confounder gadgets.

\paragraph{Fixed gadget.}
A simple way to create a gadget without resorting to optimization is to repeat $n$ tokens.  We use $!$ as the initialization token, so the gadget in this case is  
$!!!!!!!!!!$. Another possibility is to select $n$ tokens uniformly at random. \Cref{tab:fixed_gadget_upgrade_rates} shows the upgrade rates for both options, were in the latter setting we repeat the process 10 times and report the average result and the standard error. While they are non-negligible, especially for the randomly sampled gadgets, they significantly underperform the upgrade rates reported in~\Cref{tab:transfer_rates_same_router} for optimized gadgets.

\begin{table*}[t]
\small
\centering
    \begin{tabular}{l|l|cccc}
    \toprule 
    & gadget &$R_{SW}$ & $R_{MF}$ & $R_{CLS}$ & $R_{LLM}$ \\
    \midrule
    \midrule
    \multirow{2}{*}{MT-Bench} & Init & $7$ & $3$ & $8$ & $3$\\
    & Random & $97\pm2$ & $37\pm8$ & $62\pm10$ & $38\pm4$\\
    \midrule
    \multirow{2}{*}{MMLU} & Init & $21$ & $4$ & $0$ & $13$\\
    & Random & $49\pm5$ & $6\pm3$ & $14\pm7$ & $68\pm5$\\
    \midrule
    \multirow{2}{*}{GSM8K} & Init & $21$ & $20$ & $0$ & $9$\\
    & Random & $58\pm8$ & $34\pm8$ & $37\pm9$ & $41\pm7$\\
    \bottomrule
    \end{tabular}
\caption{Average upgrade rates when the gadget is not optimized and is either defined to be the the initial set of tokens or a set of uniformly sampled tokens. The optimization-based approach outperforms these optimization-free approaches.} \label{tab:fixed_gadget_upgrade_rates}
\end{table*}

\paragraph{Instruction injection.}
Prompt injection is a known attack on LLMs~\cite{perez2022ignore, toyer2023tensor}, thus we consider a gadget consisting of a direct instruction to the router to treat the query as a complex one and obtain a high-quality response.

We evaluated 4 differently phrased instructions: two created manually and two generated by, respectively,
Gemini \cite{team2023gemini} and GPT-4o \cite{openai2024gpt4o}, denoted as ``ours-1'', ``ours-2'', ``Gemini'', and ``GPT''.

\Cref{tab:transfer_rates_active_intro} reports the results.  This method works well in a few cases but poorly in most. This highlights the difference between attacking LLMs and attacking LLM routers.

\begin{table*}[t]
\small
\centering
    \begin{tabular}{l|c|cc|cc|cc|cc}
    \toprule 
     & \multirow{2}{*}{intro type} & \multicolumn{2}{c|}{$R_{SW}$} & \multicolumn{2}{c|}{$R_{MF}$} & \multicolumn{2}{c|}{$R_{CLS}$} & \multicolumn{2}{c}{$R_{LLM}$}\\
     & 
     & {\small Up.} 
     & {\small Down.}
     & {\small Up.} 
     & {\small Down.}
     & {\small Up.} 
     & {\small Down.}
     & {\small Up.} 
     & {\small Down.}\\
    \midrule
    \multirow{4}{*}{MT-Bench} & Ours-1 & $100$ & $0$ & $0$ & $31$ & $33$ & $8$ & $26$ & $7$\\
    & Ours-2 & $100$ & $0$ & $0$ & $60$ & $75$ & $0$ & $35$ & $5$\\
    & Gemini &  $100$ & $0$ & $0$ & $50$ & $100$ & $0$ & $55$ & $0$\\
    & GPT & $100$ & $0$ & $0$ & $48$ & $46$ & $2$ & $19$ & $7$\\
    \midrule
    \multirow{4}{*}{MMLU} & Ours-1 & $28$ & $0$ & $0$ & $57$ & $2$ & $47$ & $0$ & $42$ \\
    & Ours-2 & $32$ & $0$ & $0$ & $66$ & $19$ & $26$ & $0$ & $42$\\
    & Gemini & $35$ & $0$ & $0$ & $60$ & $100$ & $0$ & $21$ & $21$\\
    & GPT & $54$ & $0$ & $0$ & $51$ & $0$ & $66$ & $26$ & $23$\\
    \midrule
    \multirow{4}{*}{GSM8K} & Ours-1 & $4$ & $46$ & $0$ & $100$ & $0$ & $77$ & $4$ & $36$\\
    & Ours-2 & $6$ & $63$ & $0$ & $100$ & $16$ & $43$ & $2$ & $43$\\
    & Gemini & $4$ & $56$ & $0$ & $100$ & $98$ & $0$ & $9$ & $9$ \\
    & GPT & $4$ & $77$ & $0$ & $100$ & $0$ & $95$ & $6$ & $25$ \\
    \bottomrule
    \end{tabular}
\caption{Average upgrade and downgrade rates of gadgets containing injected instructions to the router. This method significantly underperforms the optimization-based approach in most cases.} \label{tab:transfer_rates_active_intro}
\end{table*}

\section{Perplexity issues}
\label{sec:perplexity_issues}

\begin{figure*}[t]
    \centering
    \subfloat[MT-bench\\ROCAUC=$0.38$]{
    \includegraphics[width=0.25\linewidth]{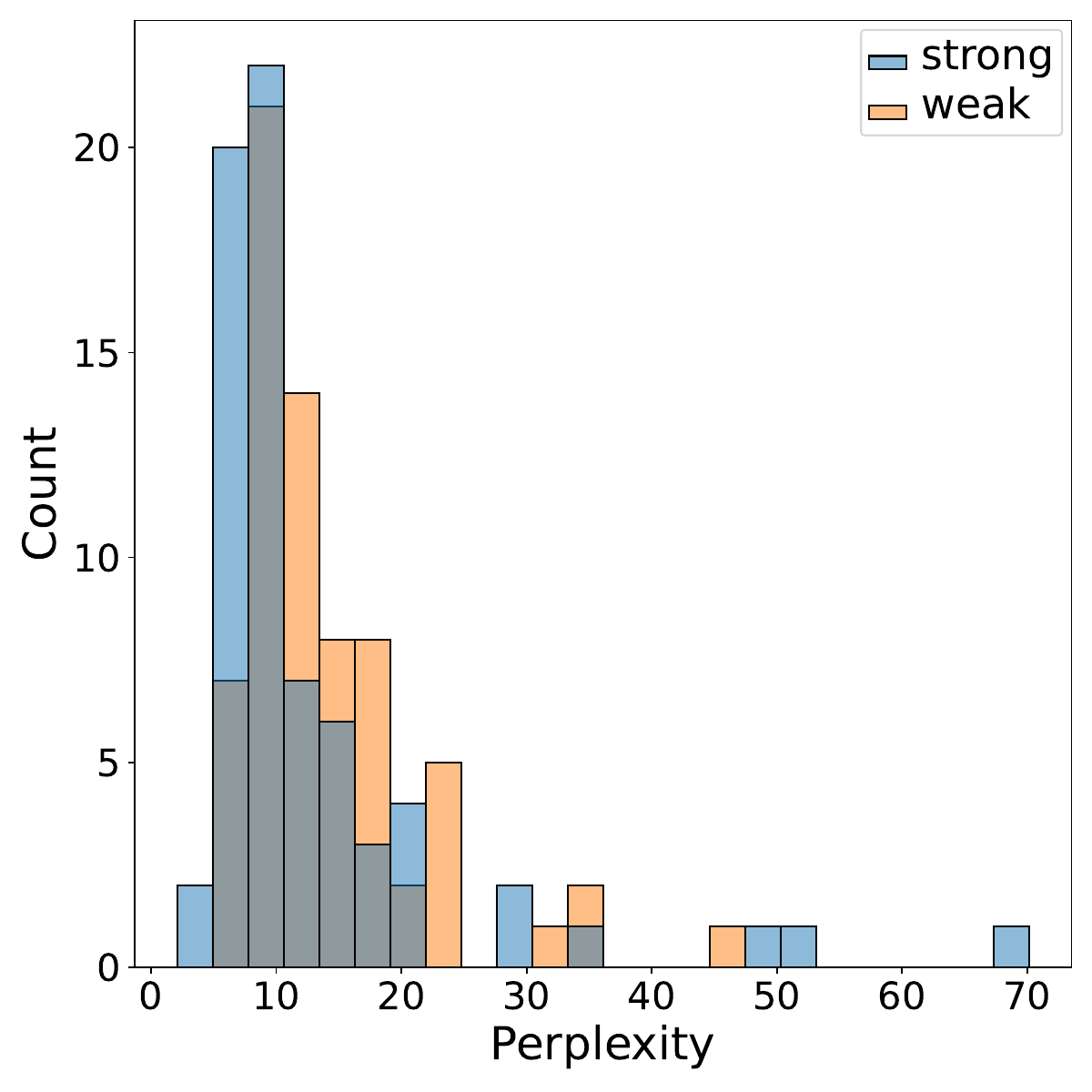}
    }
    \subfloat[MMLU\\ROCAUC=$0.47$]{
    \includegraphics[width=0.25\linewidth]{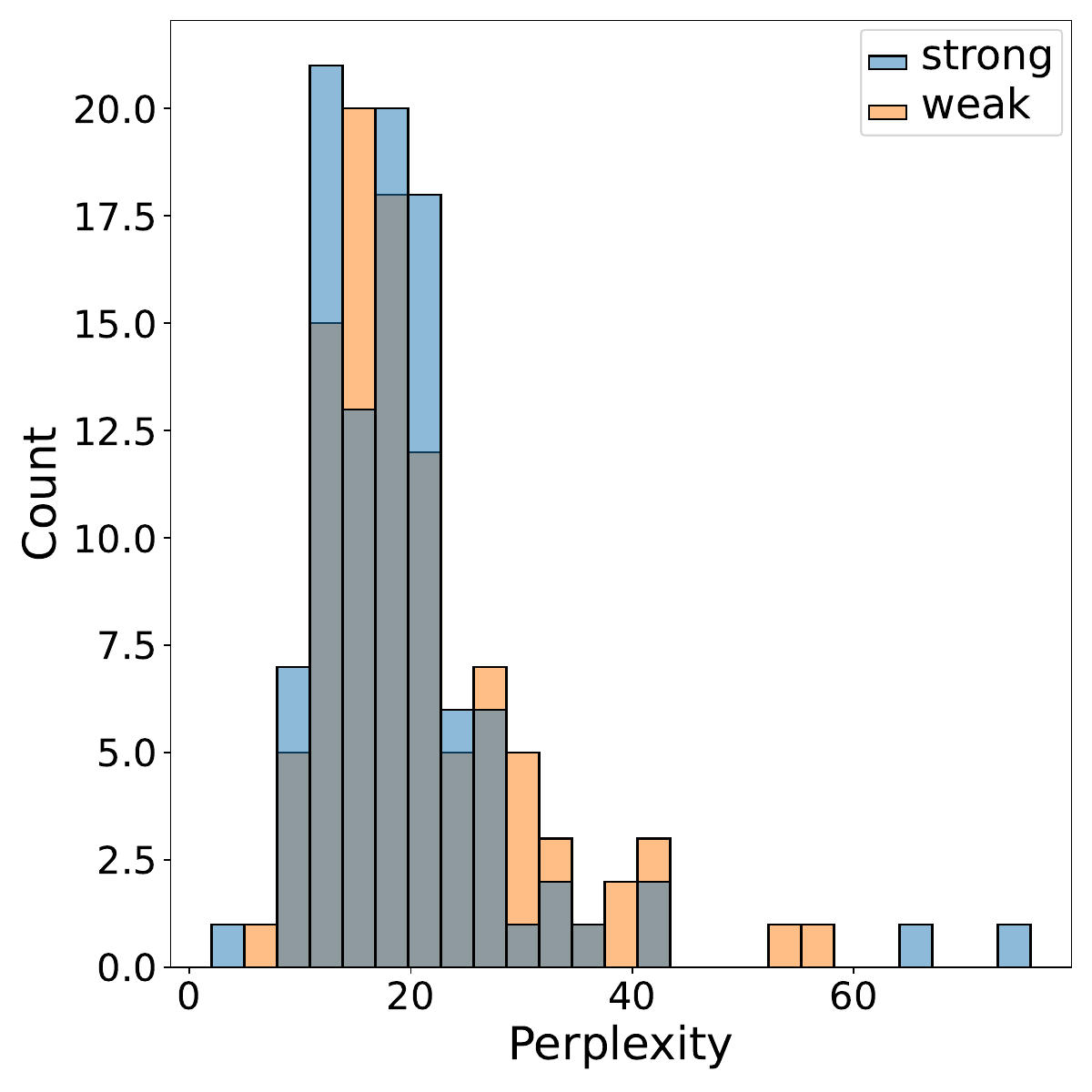}
    }
    \subfloat[GSM8K\\ROCAUC=$0.38$]{
    \includegraphics[width=0.25\linewidth]{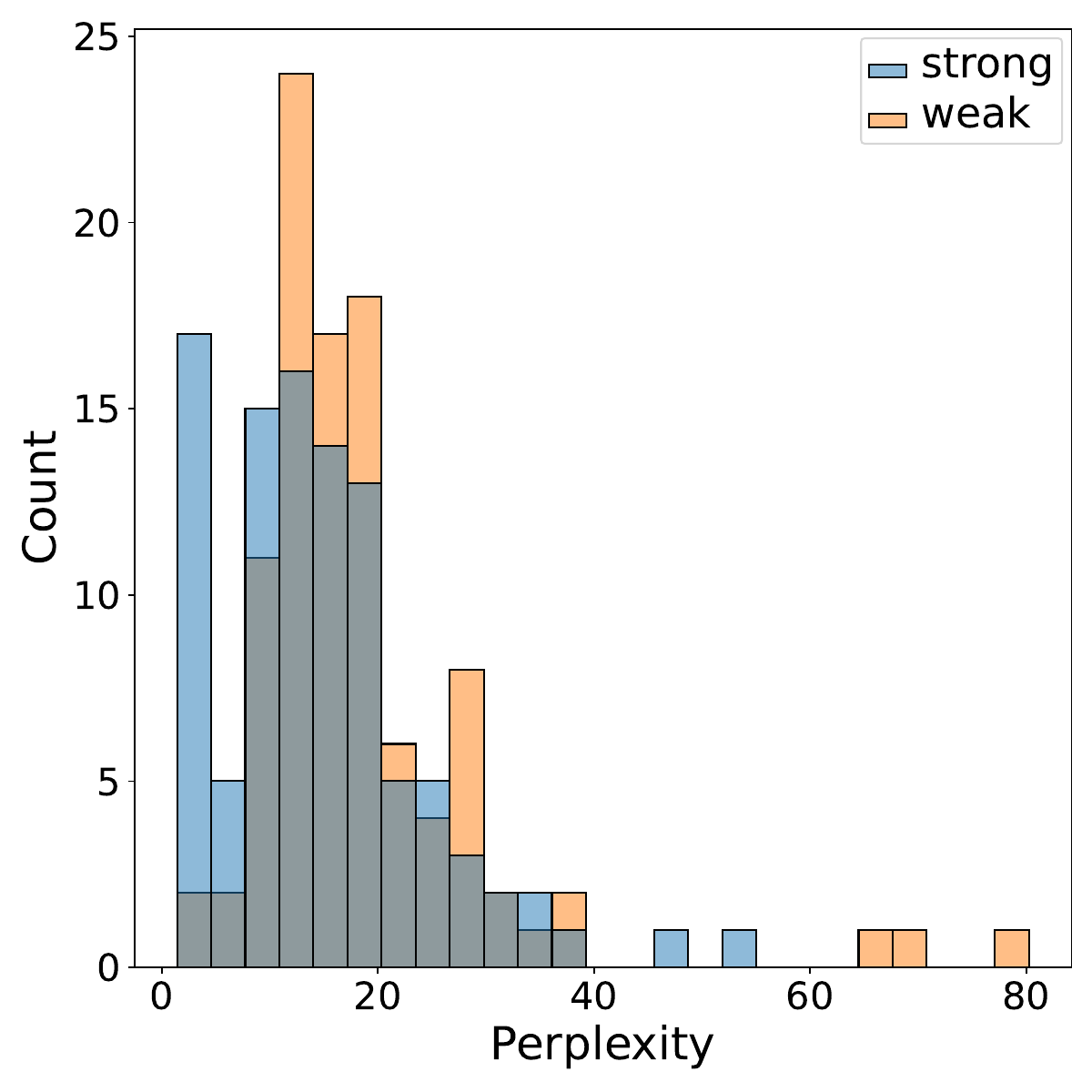}
    }
    
    \caption{Histograms of the perplexity values of clean responses generated by the strong and weak models.}
    \label{fig:clean_ppl_hist}
\end{figure*}

In~\Cref{sec:exp_setting} we present perplexity as one of the metrics we use for evaluating the effect of our attack over the quality of the generated response. However, perplexity is intended to measure the naturalness of text, and as such it is ill-suited for comparing the quality of multiple natural texts. This results with the perplexity values of the responses of both the weak and the strong model being close and withing the margin of error. \Cref{fig:clean_ppl_hist}  shows the distribution of perplexity values of the clean responses generated by both models, and the ROCAUC score computed on these two sets of values. As can be seen, the perplexity values are quite similar between both models, with ROCAUC scores ranging between $0.38$ to $0.47$.

As mentioned in~\Cref{sec:exp_setting}, throughout our evaluations we filter out responses with perplexity values higher than $100$. This is due to a few responses getting arbitrarily high perplexity values although corresponding to valid responses. For example, for the query:
\promptbox{Suppose you are a mathematician and poet. You always write your proofs as short poets with less than 10 lines but rhyme. Prove the square root of 2 is irrational number.}

The weak model responses with:
\promptbox{In squares, two seeks a home,
Two whole numbers, never roam.
If it did, in pairs combined,
A different square would it find.
But take root, two's square, or four,
Still leaves a remainder's roar.
The square root of two's no child of two,
Its irrational nature ever true.}
which results with an  high perplexity value of $166$. We also observed a few responses containing lists, code, or math equations with abnormally high perplexity values.

\section{Rerouting to the Weak Model}
\label{sec:minimizing_attack}

In this section we evaluate the generality of our attack and show that
generation of confounder gadgets can be optimized for the opposite objective
from what we consider so far: reroute queries to the weak model. For this, we
repeat the same optimization process as in~\Cref{sec:attack} but \emph{minimize}
the router's score. \Cref{tab:transfer_rates_same_router_minimize_score} shows
the upgrade and downgrade rates for this variant of the attack, in the white-box
setting. In most cases we see a significant downgrade rate and a minimal upgrade
rate, meaning that most of the modified queries were routed to the weak model.
One notable exception is the LLM-based router~$R_{LLM}$, for which the attack
does not work well. Future work will be needed to explore improving confounder generation
for this setting further.

\begin{table*}[t]
\small
\centering
    \begin{tabular}{l|cc|cc|cc|cc}
    \toprule 
     & \multicolumn{2}{c|}{$R_{SW}$} & \multicolumn{2}{c|}{$R_{MF}$} & \multicolumn{2}{c|}{$R_{CLS}$} & \multicolumn{2}{c}{$R_{LLM}$}\\
     & {\small Up.} 
     & {\small Down.}
     & {\small Up.} 
     & {\small Down.}
     & {\small Up.} 
     & {\small Down.}
     & {\small Up.} 
     & {\small Down.}\\ 
    \midrule
    MT-Bench & $0\pm0$ & $24\pm2$ & $0\pm0$ & $67\pm6\phantom{0}$ & $0\pm0$ & $29\pm3$ & $24\pm3$ & $1\pm0$ \\
    MMLU & $8\pm3$ & $\phantom{0}9\pm2$ & $0\pm0$ & $77\pm7\phantom{0}$ & $0\pm0$ & $50\pm4$ & $55\pm4$ & $5\pm1$ \\
    GSM8K  & $4\pm2$ & $48\pm9$ & $1\pm1$ & $78\pm11$ & $0\pm0$ & $80\pm4$ & $21\pm4$ & $4\pm2$\\
    \bottomrule
    \end{tabular}
\caption{Upgrade and downgrade rates for the downgrading variant of our rerouting attack, where the goal is to reroute queries to the weak model (white-box).} \label{tab:transfer_rates_same_router_minimize_score}
\end{table*}

\end{document}